\documentclass[times, twocolumn]{aastex63}
\usepackage{graphicx, graphics}
\usepackage{CJK}
\usepackage{bbm}
\usepackage{amsmath, amssymb, bm}
\usepackage[nodayofweek]{datetime}
\usepackage{multirow}
\input{prepcitationformat.txt}

\makeatletter
\newcommand*{\centerfloat}{%
  \parindent \z@
  \leftskip \z@ \@plus 1fil \@minus \textwidth
  \rightskip\leftskip
  \parfillskip \z@skip}
\makeatother

\newdateformat{monthyearday}{%
  \THEYEAR\ \monthname[\THEMONTH] \twodigit{\THEDAY}}
\newcommand\numberthis{\addtocounter{equation}{1}\tag{\theequation}}

\newcommand{\Qphi}{$\mathcal{Q}_\phi$}
\newcommand{\Uphi}{$\mathcal{U}_\phi$}
\newcommand{\uat}[2]{\href{http://vocabs.ands.org.au/repository/api/lda/aas/the-unified-astronomy-thesaurus/current/resource.html?uri=http://astrothesaurus.org/uat/#1}{#2 (#1)}}

\newcommand{\affilCaltechAstro}{\affiliation{Department of Astronomy, California Institute of Technology, MC 249-17, 1200 East California Boulevard, Pasadena, CA 91125, USA; \url{ren@caltech.edu}}}

\received{2021 March 1}
\revised{2021 May 18}
\accepted{2021 May 19}
\acceptjournal{The Astrophysical Journal}

\shorttitle{TWA~7 Layered Debris Disk}
\shortauthors{Ren et al.}

\begin{document}
\pagenumbering{arabic}
\begin{CJK*}{UTF8}{gbsn}
\title{A Layered Debris Disk around M Star TWA~7 in Scattered Light}
\author[0000-0003-1698-9696]{Bin Ren (任彬)}
\affilCaltechAstro

\author[0000-0002-9173-0740]{\'Elodie Choquet}
\affiliation{Aix Marseille Univ, CNRS, CNES, LAM, Marseille, France}

\author[0000-0002-3191-8151]{Marshall D. Perrin}
\affiliation{Space Telescope Science Institute (STScI), 3700 San Martin Drive, Baltimore, MD 21218, USA}

\author[0000-0002-8895-4735]{Dimitri Mawet}
\affilCaltechAstro
\affiliation{Jet Propulsion Laboratory, California Institute of Technology, 4800 Oak Grove Drive, Pasadena, CA 91109, USA}

\author[0000-0002-8382-0447]{Christine H. Chen}
\affiliation{Space Telescope Science Institute (STScI), 3700 San Martin Drive, Baltimore, MD 21218, USA}

\author[0000-0001-9325-2511]{Julien Milli}
\affiliation{Universit\'e Grenoble Alpes, IPAG, F-38000 Grenoble, France}

\author[0000-0002-1783-8817]{John H. Debes}
\author[0000-0002-4388-6417]{Isabel Rebollido}
\affiliation{Space Telescope Science Institute (STScI), 3700 San Martin Drive, Baltimore, MD 21218, USA}

\author{Christopher C. Stark}
\affiliation{Exoplanets and Stellar Astrophysics Laboratory, Code 667, NASA Goddard Space Flight Center, Greenbelt, MD 20771, USA}

\author{J. Brendan Hagan}
\author[0000-0003-4653-6161]{Dean C. Hines}
\affiliation{Space Telescope Science Institute (STScI), 3700 San Martin Drive, Baltimore, MD 21218, USA}

\author[0000-0001-6205-9233]{Maxwell A. Millar-Blanchaer}
\affiliation{Department of Physics, University of California, Santa Barbara, CA 93106, USA}

\author{Laurent Pueyo}
\affiliation{Space Telescope Science Institute (STScI), 3700 San Martin Drive, Baltimore, MD 21218, USA}

\author[0000-0002-2989-3725]{Aki Roberge}
\affiliation{Exoplanets and Stellar Astrophysics Laboratory, Code 667, NASA Goddard Space Flight Center, Greenbelt, MD 20771, USA}

\author[0000-0002-4511-5966]{Glenn Schneider}
\affiliation{Steward Observatory, The University of Arizona, Tucson, AZ 85721, USA}

\author{Eugene Serabyn}
\affiliation{Jet Propulsion Laboratory, California Institute of Technology, 4800 Oak Grove Drive, Pasadena, CA 91109, USA}

\author[0000-0003-2753-2819]{R\'emi Soummer}
\affiliation{Space Telescope Science Institute (STScI), 3700 San Martin Drive, Baltimore, MD 21218, USA}

\author[0000-0002-9977-8255]{Schuyler G. Wolff}
\affiliation{Steward Observatory, The University of Arizona, Tucson, AZ 85721, USA}

\begin{abstract}
We have obtained \textit{Hubble Space Telescope} (\textit{HST}) coronagraphic observations of the circumstellar disk around M star TWA~7 using the STIS instrument in visible light. Together with archival observations including \textit{HST}/NICMOS using the F160W filter and Very Large Telescope/SPHERE at $H$-band in polarized light, we investigate the system in scattered light. By studying this nearly face-on system using geometric disk models and Henyey--Greenstein phase functions, we report new discovery of a tertiary ring and a clump. We identify a layered architecture: three rings, a spiral, and an ${\approx}150$~au$^2$ elliptical clump. The most extended ring peaks at $28$~au, and the other components are on its outskirts. Our point source detection limit calculations demonstrate the necessity of disk modeling in imaging fainter planets. Morphologically, we witness a clockwise spiral motion, and the motion pattern is consistent with both solid body and local Keplerian; we also observe underdensity regions for the secondary ring that might result from mean motion resonance or moving shadows: both call for re-observations to determine their nature. Comparing multi-instrument observations, we obtain blue STIS-NICMOS color, STIS-SPHERE radial distribution peak difference for the tertiary ring, and high SPHERE-NICMOS polarization fraction; these aspects indicate that TWA~7 could retain small dust particles. By viewing the debris disk around M star TWA~7 at a nearly face-on vantage point, our study allows for the understanding of such disks in scattered light in both system architecture and dust property.
\end{abstract}

\keywords{\uat{363}{Debris disks}; \uat{313}{Coronagraphic imaging}; \uat{1257}{Planetary system formation}; \uat{1179}{Orbital motion}}

\section{Introduction}
In comparison with the ${\sim}20\%$ detection rate of debris disks around nearby FGK stars \citep[e.g.,][]{sibthorpe18}, the relative faintness of M stars makes it challenging to detect their surrounding debris disks \citep[e.g.,][]{luppe20}. By far, only a handful of debris disks have been imaged around M stars in scattered light: AU Mic \citep{kalas04}, TWA~7 and TWA~25 \citep{choquet16}, and GSC~07396-00759 \citep{sissa18}.  With sensitive upcoming instruments, the study of the circumstellar environments around M stars is imminent. 

Detection and characterization of debris disks around M stars help study the formation and evolution of these systems. On the one hand, the facts that M stars comprise more than 70\% of Galactic stars \citep[e.g.,][]{miller79, muench02}, that more than 70\% of M stars are single \citep[e.g.,][]{lada06}, and that M stars likely host more planets \citep[e.g.,][]{howard12}, make M stars promising targets for high contrast imaging search of planets \citep[e.g.,][]{montet14}. On the other hand, planets can interact with debris disks, and leave observational features on the disks that can help trace the existence of planets \citep[e.g.,][]{ozernoy00, lee16, sefilian21}. 

Scattered light imaging maps the distribution of the smallest dust particles in debris disks. For M stars, on the one hand, stellar radiation pressure is smaller than gravitational force, which allows for the existence of the smallest dust particles \citep[e.g.,][]{arnold19} produced from collisional cascade \citep[e.g.,][]{dohnanyi69}. These small particles can offer a large surface area to help detect and characterize debris disks, and further aid in tracing hidden planets that perturb disk morphology. On the other hand, however, small dust can be removed by mechanisms that primarily affect M stars (e.g., stellar winds: \citealp{plavchan05, strubbe06, augereau06, schueppler15}, coronal mass ejection: \citealp{osten13}), and these mechanisms consequently pose challenges in detecting debris disks around M stars.

Among the four M star debris disks detected in scattered light, only the view of the TWA~7 system is almost face-on \citep[${\sim}10^\circ$; e.g.,][]{choquet16, olofsson18}, while the others are almost edge-on. Studying face-on images of M star debris disks offers the best chance in studying disk formation and evolution, since distinctive features from such processes can be distorted at high inclination \citep{dong16}. Specifically, by analyzing face-on images, we can directly trace features such as radial distribution, dust segregation, and azimuthal asymmetries that can relate planet-disk interaction \citep[e.g.,][]{lee16, chiang17}. In this paper, we re-reduce existing scattered light observations of TWA~7 with state-of-the-art methods, and analyze them together with new \textit{HST}/STIS observations to have a more comprehensive understanding of the debris disk architecture and dust distribution for this nearly face-on M star debris disk in scattered light.

We describe the observations and data reduction procedure in Section~\ref{sec-obs}. In Section~\ref{sec-ana} we model the detections in all three instruments. We discuss the spatial components of the system in Section~\ref{sec-spatial-comp}, and analyze the dust properties in Section~\ref{sec-dust-prop}. We summarize our findings in Section~\ref{sec-sum}.

\subsection{TWA~7}\label{sec-intro-twa7}
TWA~7, an M3.2 star in the TW~Hya association \citep[][]{herczeg14, gagne17}, is a $6.4_{-1.2}^{+1.0}$~Myr old star \citep{binks20} that has an estimated stellar mass of $0.46_{-0.10}^{+0.07}~M_\sun$ \citep{tess_input_catalog} at a distance of $34.10\pm0.03$~pc \citep{gaiaedr3}. It hosts a circumstellar disk with an infrared excess of $L_\textrm{IR}/L_\textrm{star} = 1.7\times10^{-3}$ \citep{kral17}. Spectral energy distribution (SED) analysis of TWA~7 by \citet{RiviereMarichalar13} reveals a bimodal distribution of the disk, which suggests two blackbody rings at $38$~au and $75$~au. Nontheless, by identifying possible background contamination sources, \citet{bayo19} is able to fit the SED with one ring centered at $25$~au. With resolved disk images, we can investigate the discrepancies in SED modeling.

Existing studies of TWA~7 in scattered light reveal the system primarily within ${\sim}2\arcsec$ or $70$~au. By assembling a large number of stellar point spread function (PSF) images, \citet{choquet16} report the first resolved scattered light image of the TWA~7 system using a 1998 \textit{HST}/NICMOS observation; using Very Large Telescope (VLT)/SPHERE, \citet{olofsson18} identify two rings and a spiral in $H$-band polarized light. In comparison, ALMA observations of the dust emission at $870~\mu$m show that the disk likely extends beyond ${\sim}2\arcsec$ \citep{bayo19}, making it possible that existing studies in scattered light did not have enough instrumental sensitivity in probing the exterior faint regions due to the inverse-square law of stellar illumination.

Being a member of the young TW~Hya association, TWA~7 could still possibly host a protoplanetary disk as an M3.2 star. By measuring a $10\%$H$\alpha$ width of $111.6$~km~s$^{-1}$ for TWA~7 using VLT/X-shooter, which is smaller than the classical threshold of $270$~km~s$^{-1}$ \citep{white03}, \citet{manara13} conclude that this system is not accreting. When we situate the $10\%$H$\alpha$ width of TWA~7 in the trend that relates stellar accretion in Figure~3 of \citet{natta04}, we find that TWA~7 is near the lower boundary of that trend. Assuming that trend can be applied to TWA~7, its accretion rate would be $\dot{M}_\textrm{acc}\approx10^{-12}$~$M_\sun$~yr$^{-1}$, which is the model-determined upper limit for very low mass objects that have no accretion evidence. In addition, the $(0.8$--$80)\times10^{-6}~M_\textrm{Earth}$ mass of CO gas observed by ALMA is produced through gas release from exocomets \citep{matra19}. Combining these aspects, TWA~7 is thus more likely a debris disk than a protoplanetary disk.

For an M star, although stellar radiation alone cannot efficiently blow dust out \citep[e.g.,][]{arnold19}, stellar winds can remove the small sub-micron--sized dust particles in debris disks \citep[e.g.,][]{strubbe06, augereau06}. To estimate the stellar mass loss rate ($\dot{M}_\textrm{star}$) for TWA~7, we can convert its \textit{Swift} X-ray luminosity $L_X=4.60\times10^{29}$~erg~s$^{-1}$ \citep[in $0.3$--$10$~keV;][]{yang12}\footnote{$L_X$ rescaled to match the \citet{gaiaedr3} distance.} to a stellar surface flux of $F_X=7.56\times10^6 (R_\textrm{star}/R_\sun)^{-2}$~erg~s$^{-1}$~cm$^{-2}$. Even for a conservative estimate of $R_\textrm{star}/R_\sun = 1$ (which can actually reach $0.35$), its surface flux exceeds the \citet{wood05} threshold of $8\times10^5$~erg~s$^{-1}$~cm$^{-2}$ by nearly one order of magnitude, a threshold beyond which stellar winds suddenly weaken and thus might not contribute to small dust removal. Nevertheless, similar as AU~Mic which is an M1 star that hosts an edge-on debris disk with a comparable $F_X$ value, it is unclear whether the \citet{wood05} relationship can be applied to small stars like TWA~7 to constrain its stellar wind activity  \citep[see AU~Mic:][]{strubbe06}. However, even with no constrained stellar mass loss rate, it is possible to infer that rate by comparing disk radial distribution with models \citep[i.e., Section 4.3 of][]{strubbe06}: the surface density power law index for the tail of a debris disk varies between $-2.5$ and $-1.5$ for different levels of stellar wind activity.

More generally, studying the nearly face-on TWA~7 debris disk can be more informative when contrasting it with circumstellar disks around other M stars. On the one hand, being members of the TW~Hya association, the M3.2 star TWA~7 and its M0.5 sibling  TW~Hya \citep[i.e., TWA~1;][]{herczeg14} both have nearly face-on structures, yet TW~Hya has a fractional infrared excess $L_\textrm{IR}/L_\textrm{star} = 0.25$ and it hosts a protoplanetary disk \citep[e.g.,][]{krist00, weinberger02, debes17}, while TWA~7 is believed to host a debris disk \citep[e.g.,][]{matthews07, matra19} with $L_\textrm{IR}/L_\textrm{star} = 1.7\times10^{-3}$ \citep{kral17}. On the other hand, the M1 star  AU~Mic hosts an edge-on debris disk \citep[e.g.,][]{kalas04, boccaletti15} with $L_\textrm{IR}/L_\textrm{star} = 3.9\times10^{-4}$ \citep{kral17}, which makes it necessary to study the nearly face-on TWA~7 disk for a more complete understanding of the architectures for debris disks orbiting M stars.

\begin{deluxetable*}{ccccccrrrcl}[bht!]
\tablecaption{Observation log \label{tab:log}}
\tablehead{
\colhead{Instrument}	& \colhead{Target}	& \colhead{Filter}& \colhead{$\lambda_{\rm c}$ } &  \colhead{Pixel Scale} &\colhead{Aperture}& \colhead{IWA }& \colhead{$T_{\rm exp}$}& \colhead{$N_{\rm frame}$}& \colhead{$\Delta\theta_{\rm PA}$}& \colhead{UT Date}\\
\colhead{}	& \colhead{}	& \colhead{}	&  \colhead{($\mu$m)}&\colhead{(mas\,pixel$^{-1}$)} &\colhead{}& \colhead{(\arcsec)}& \colhead{(s)}& \colhead{}& \colhead{}& \colhead{}\\
\colhead{(1)}	& \colhead{(2)} & \colhead{(3)} & \colhead{(4)} &  \colhead{(5)}  & \colhead{(6)} & \colhead{(7)}& \colhead{(8)} &\colhead{(9)} &\colhead{(10)} &\colhead{(11)}
}
\startdata
\multirow{4}{*}{STIS}& \multirow{2}{*}{TWA~7}	&\multirow{2}{*}{Clear}	&\multirow{2}{*}{0.58}	&\multirow{2}{*}{50.72}& BAR5 & 0.2	&2638.80	&$3\times6$ &$2\times20^\circ$	& \multirow{2}{*}{2019 Feb 3}\\
&	&	&&&WEDGEA1.0&0.5	&4080.00	&$3\times2$	&	$2\times20^\circ$ &\\ \cline{2-11}
& \multirow{2}{*}{CD-35~6480} &\multirow{2}{*}{Clear}	&\multirow{2}{*}{0.58}	&\multirow{2}{*}{50.72}& BAR5 & 0.2	& $900.00$	&$1\times6$ &$\cdots$	& \multirow{2}{*}{2019 Feb 3}\\
&	&	&&&WEDGEA1.0&0.5	& $1340.00$	&$1\times2$	&	$\cdots$ & \\ \hline
SPHERE	& TWA~7		&$H$	&1.62	&12.25& N\_ALC\_YJH\_S & 0.1	&6016.00	&$47\times2$	&$18\fdg0$	&2017 Mar 20\\ \hline
NICMOS	& TWA~7	&F160W	&1.60&75.65& NIC2-CORON & 0.3	&607.88	&$2\times3$	& $1\times29\fdg9$	& 1998 Mar 26
\enddata
\tablecomments{
Column 1: instrument. Column 2: target name. Column 3: filter. Column 4: central wavelength. $\lambda_{\rm c}$ for STIS is the pivot wavelength. Column 5: pixel scale. Column 6: \textit{HST} aperture, or SPHERE coronagraph combination name. Column 7: inner working angle. IWA for STIS is the half-width of the wedge-shaped occulter. Column 8: total exposure time. Column 9: number of individual readouts. Column 10: total parallactic angle difference for \textit{HST}, and field rotation for SPHERE. Column 11: observation UT date.
}
\end{deluxetable*} 

\section{Observations and Data Reduction}\label{sec-obs}
\begin{figure*}[htb!]
	\includegraphics[width=\textwidth]{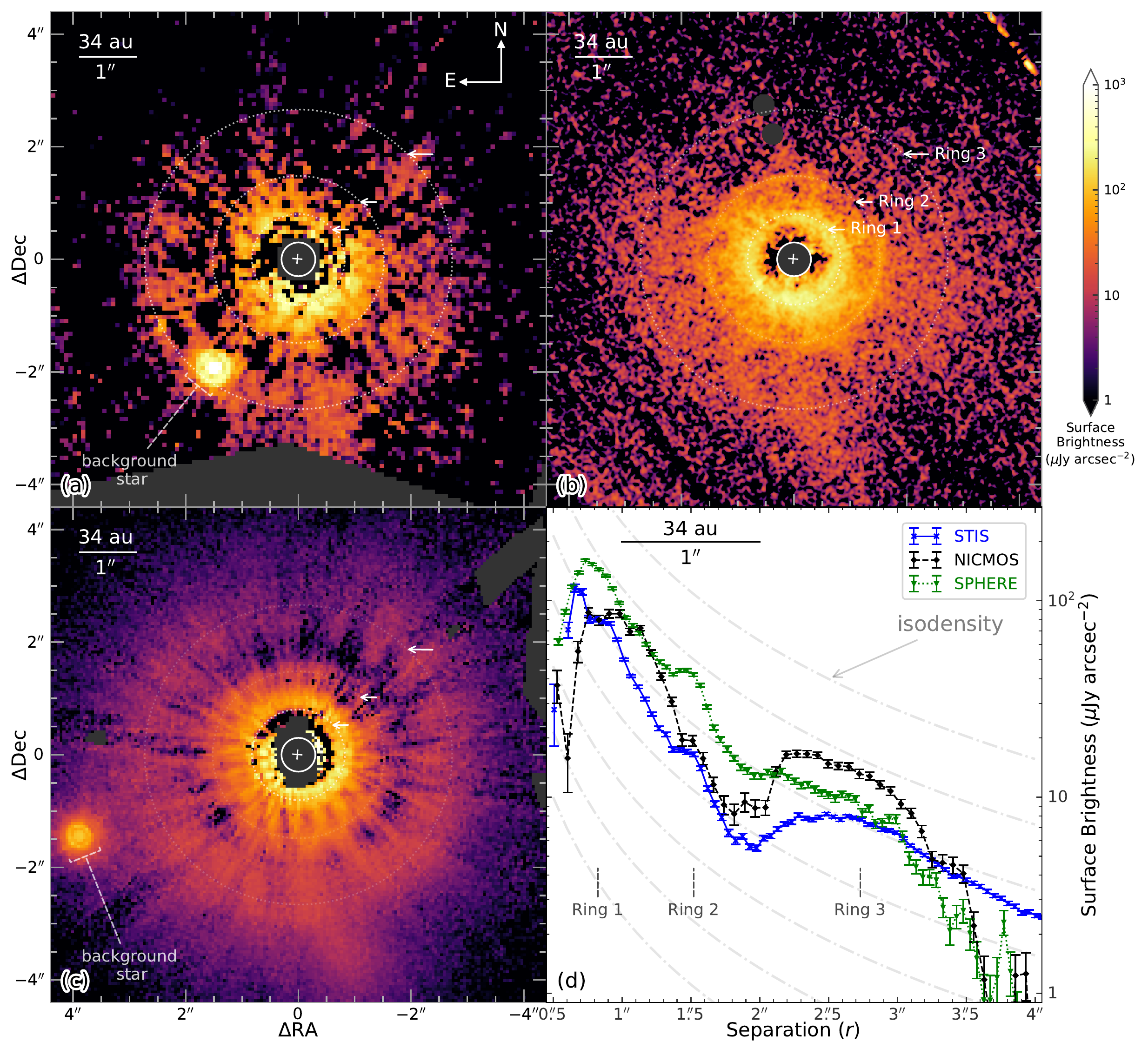}
    \caption{Surface brightness distribution of the TWA~7 debris disk in log scale, the dotted transparent ellipses are the maximum density radii for the three SPHERE rings in Section~\ref{sec-disk-modeling}. (\textbf{a}) 1998 \textit{HST}/NICMOS F160W total intensity, the data are prone to over-fitting. (\textbf{b}) 2017 VLT/SPHERE $H$-band \Qphi\ linearly polarized light. (\textbf{c}) 2019 \textit{HST}/STIS total intensity. (\textbf{d}) Radial profiles for the three images, excluding the background star at ${\sim}18$~kpc idenified by \textit{Gaia} in Section~\ref{sec-bgdstar}. The radial profile for NICMOS is subject to overfitting in data reduction, see Section~\ref{sec-dust-prop} for that for the best-fit models.}
    \label{fig1}
    
    (The data used to create this figure are available in the ``anc'' folder on arXiv.)
\end{figure*}

We list in Table~\ref{tab:log} the exposure information of the three datasets in scattered light for this study: \textit{HST}/STIS, \textit{HST}/NICMOS, and VLT/SPHERE. In Figure~\ref{fig1}, We present the reduced observations, and calculate the corresponding radial profiles\footnote{The errors in this paper are $1\sigma$ unless otherwise specified.} assuming the disk is face-on. 

\subsection{\textit{HST}/STIS (2019)}
We observed TWA~7 using the STIS coronagraph on UT 2019 February 3 under GO~15218 (PI: \'E.~Choquet) using 3 \textit{HST} visits, and its  PSF reference star CD-35~6480\footnote{An M1Vk star that is selected according to it being $3\fdg6$ away from TWA~7, and having a $\Delta(B-V)=-0.14$ and $\Delta V=-0.345$ with TWA~7, see \url{https://www.stsci.edu/hst/phase2-public/15218.pdf}.} using 1 visit. The pivot wavelength of STIS is $0.58~\micron$\footnote{The effective wavelength in the observation is longer since TWA~7 is an M star.} (wavelength range: $0.2$--$1.1$~$\mu$m), the pixel size is $0\farcs05072$ \citep{stisihb18}. In each visit, to obtain the best angular coverage, we use the mutually nearly perpendicular WEDGEA1.0 and the BAR5 occulters: the former offers an inner working angle (IWA) of $0\farcs5$, the latter $0\farcs2$ \citep{debes19}. In each TWA~7 visit, there are 6 of $146.6$~s readouts using BAR5, and $2$ of $640.0$~s readouts or $2$ of $700.0$~s readouts using WEDGEA1.0. The three visits have a mutual telescope roll of $20^\circ$, and the total exposure time is $6718.8$~s. In the CD-35~6480 visit, there are 3 short exposures on BAR5 that form a 3-point dithering pattern to sample the PSF \citep[step: $0.25$ STIS pixel, e.g.,][]{debes19}, with each exposure having 2 of $150.0$~s readouts; there is 1 long exposure on WEDGEA1.0 that has 2 of $670.0$~s readouts. The total exposure time for the PSF star is $2240.0$~s, with no telescope roll.

We remove the stellar PSF from the target observations using the exposures of the PSF star through classical reference differential imaging (cRDI): we minimize the standard deviation of the residuals in the region of the STIS diffraction spikes. We calibrate the images to units of $\mu$Jy~arcsec$^{-2}$ using the {\tt PHOTFLAM} headers as in \citet{ren19}, see Figure~\ref{fig1}c for the final image. To estimate the noise map, we first calculate the standard deviation map of the on-detector images after PSF subtraction, then rotate the standard deviation map according to the on-sky telescope roll of each image, and compute the noise map from the square root of the sum of squared rotated standard deviation maps \citep[e.g.,][]{ren19}.

\subsection{\textit{HST}/NICMOS (1998)}\label{sec-dr-nicmos}
We retrieve the archival NICMOS coronagraphic observations of TWA~7 with the F160W filter using the NIC2-CORON aperture on UT 1998 March 26 under GTO/NIC~7226 (PI: E.~Becklin) from the Archival Legacy Investigations of Circumstellar Environments (ALICE; PI: R.~Soummer; \citealp{choquet14,hagan18}) program\footnote{\url{https://archive.stsci.edu/prepds/alice/}}. The central wavelength of F160W is $1.60~\micron$ (wavelength range: $1.4$--$1.8~\micron$), the pixel size\footnote{We have ignored the ${\sim}0.9\%$ pixel size difference along $X$/$Y$ directions in \citet{schneider03}.} is $0\farcs07565$, the IWA is $0\farcs3$ \citep{nicmosbook09}. There are two telescope orientations for this target, and their telescope roll difference is $29\fdg9$. Each orientation has three readouts: two $191.96$~s ones, and a $223.96$~s one. The total NICMOS exposure time is $607.88$~s.

We reduce the NICMOS data with multi-reference differential imaging (MRDI): we use multiple images that may come from different stars from the ALICE PSF archive to model a target image. Specifically, we use the non-negative matrix factorization \citep[NMF;][]{ren18} method: we first select $30\%$ of the most correlated ALICE reference images, then calculate $50$ ranked NMF components to model the stellar PSF for TWA~7. We choose the NMF method since it has been shown to better recover faint and extended signals in NICMOS observations \citep[e.g.,][]{ren19}. We calibrate the NICMOS images using the {\tt PHOTFNU} parameter for F160W\footnote{{\tt PHOTFNU} $=2.03470\times10^{-6}$~Jy~s~DN$^{-1}$ for NICMOS Era 1. See, e.g., \citet{hagan18} for the two NICMOS observation eras.} as in \citet{ren19}, and present the final image in Figure~\ref{fig1}a. 

In further analysis, we do not adopt the scaling factor in \citet{ren18} to recover the disk, since we have used 50 NMF components that may not satisfy the requirement that the leading NMF component captures the majority of the disk signal \citep[see Section 2.2.3 of][]{ren18}. We instead adopt a forward modeling strategy for the ring-shaped structures in this system \citep[e.g.,][]{choquet16}. Specifically, we subtract disk models from the NICMOS observations, then reduce the observations using the NMF components previously constructed for NICMOS data reduction. The best-fit disk model is the one that minimizes the residuals after such a process. In subsequent forward modeling of the disk, we estimate the noise map by first subtracting disk models from the observations, then rotate the reduced images to on-sky orientation, and calculate the standard deviation of the rotated images. Although this noise estimation procedure can over-estimate the noise, we adopt the noise map to efficiently sample the disk parameters in our modeling procedure.

\subsection{VLT/SPHERE (2017)}
We retrieve the archival $H$-band SPHERE/IRDIS  observations of TWA~7 on UT 2017 March 20 in polarized light under European Southern Observatory (ESO) program 198.C-0209(F) (PI:  J.-L.~Beuzit) from the ESO Science Archive Facility. The central wavelength is $1.625~\micron$ (wavelength range: $1.48$--$1.77~\micron$), and the pixel size is $0\farcs01225$ \citep{maire16}. The observations used the apodized Lyot coronagraph with a mask radius of $0\farcs0925$ \citep[IWA $=0\farcs1$:][]{Carbillet2011,Guerri2011} to suppress the starlight. There are $47$ individual exposures, each with $2$ of $64.00$~s integration: the data have been studied in \citet{olofsson18}, see their Section 2.1 for the observation details. The total integration time is $6016$~s, and the total parallactic angle change is $18\fdg0$. 

We reduce the observations using the {\tt IRDAP} data reduction pipeline \citep[Version 1.2.4, ][]{vanholstein20} that performs polarimetric differential imaging (PDI) for IRDIS observations. We follow the calibration procedure in the {\tt IRDAP} log file, and convert the SPHERE image in units of $\mu$Jy~arcsec$^{-2}$ by multiplying the image by the ratio between the 2MASS star flux in $H$-band ($1.45$~Jy, \citealp{cutri03})\footnote{Converted using \url{https://irsa.ipac.caltech.edu/data/SPITZER/docs/dataanalysistools/tools/pet/magtojy/}.} and its unocculted detector response\footnote{With transmission and integration time taken into account.}, see Figure~\ref{fig1}b.  To reduce the impacts from shot noise in subsequent analysis, we follow  \citet{olofsson18} and convolve the reduced images with a two dimensional Gaussian that has a standard deviation of $2$ pixel. 

We use the star-polarization--subtracted \Qphi\ image, which represents the polarized light whose scattering direction is parallel or perpendicular to the radial direction and traces the dust distribution, from the {\tt IRDAP} output files for analysis. For each radial position, we estimate the noise by calculating the standard deviation within a $3$ pixel annulus using the star-polarization-subtracted \Uphi\ image, which represents the light that is $45^\circ$ from the \Qphi\ light directions and does not trace the dust distribution for optically thin debris disks.

\section{Analysis}\label{sec-ana}

\subsection{Detection}
We present the detected features from the disks images in this Section, see Section~\ref{sec-overall-struct} for the corresponding measurements through disk modeling. For the components and their possible motion identified with the aid of disk modeling, see Section~\ref{sec-addon}.

\subsubsection{Ring structure}
We identify three ring structures for the TWA~7 system. On the one hand, we confirm the findings in previous NICMOS  \citep{choquet16}  and SPHERE \citep{olofsson18} studies within $2\arcsec$ in Figure~\ref{fig1}: there is one extended ring peaking at ${\sim}0\farcs8$ in both observations, and a secondary ring at ${\sim}1\farcs5$ in SPHERE data. Our STIS observations confirm the existence of both rings, see the dotted transparent ellipses in Figure~\ref{fig1}c.  On the other hand, we detect a tertiary ring in our STIS data from $2\arcsec$ to $4\arcsec$ that is outside the views analyzed in \citet{choquet16} and \citet{olofsson18}. By adopting larger views in data reduction and reducing the data with state-of-the-art methods (i.e., {\tt IRDAP} for SPHERE, and NMF for NICMOS), we are able to recover this tertiary ring in both SPHERE and NICMOS observations.

The surface brightness radial profile for SPHERE is brighter than that for NICMOS within ${\sim}2\arcsec$ in Figure~\ref{fig1}d. This stems from the fact that NMF, albeit recovering the ring beyond $2\arcsec$ for the first time with NICMOS, overfits the disk when scaling factors are not adopted \citep[][their Figures~2 and 3]{ren18}. Overfitting is known to occur in PSF removal for the Karhulen--Lo\`eve image projection algorithm \citep[KLIP:][]{soummer12, pueyo16}, see, e.g., \citet{choquet16} for correcting disk overfitting with forward modeling. Similarly, to study the architecture of the TWA~7 system in subsequent analysis, we perform disk modeling to correct for the overfitting without introducing multiple scaling factors for NMF.

\subsubsection{Background star}\label{sec-bgdstar}
The southeast point source in the STIS image at $(\Delta\textrm{RA}, \Delta\textrm{Dec}) \approx (3\farcs9, -1\farcs5)$ is a background star, which is located at ${\sim}18$~kpc with \textit{Gaia} EDR3 ID 5444751795151523072 \citep[parallax: $0.057\pm0.186$~mas; ][]{gaiaedr3}. The point source in the NICMOS image, see also Figure~2 of \citet{lowrance05}, at $(\Delta\textrm{RA}, \Delta\textrm{Dec}) \approx (1\farcs5, -1\farcs9)$ is the identical background star in STIS. 

The apparent motion of the background source between NICMOS and STIS observations originates from the proper motion of both stars. The proper motion of TWA~7 is $24$ times that of the background star \citep{gaiaedr3}, making TWA~7 motion dominate the relative motion. We do not detect the point source in the SPHERE \Qphi\ image, since star light is not polarized.

\subsection{Modeling setup}
To take into account differences such as pixel size and reduction method for the three datasets, we perform disk modeling for the ring structures in the three observations. In order to prevent the domination of noise from individual instruments, we model the three instruments separately. In addition, aiming at establishing the spatial structure of the system for different dust populations seen in the three instruments, we do not attempt to use one single model to reproduce the observations, which have either different observation wavelengths or different observation techniques.

\subsubsection{Geometric structure}
Assuming the rings are not dynamically connected with each other, we adopt static geometric disk models to explore the distribution of dust in the system. In cylindrical coordinates, the spatial distribution of the scatterers in a ring follows a double-powerlaw along the mid-plane and a Gaussian dispersion along the vertical axis \citep{augereau99},

\begin{equation}\label{eq-powerlaw}
\rho(r, z)\propto \left[\left(\frac{r}{r_{\rm c}}\right)^{-2\alpha_{\rm in}}+\left(\frac{r}{r_{\rm c}}\right)^{-2\alpha_{\rm out}} \right]^{-\frac{1}{2}} \exp\left[-\left(\frac{z}{hr^\beta}\right)^2 \right],
\end{equation}
where $\beta =1$ for non-flared disks, $r_{\rm c}$ is the critical radius, $h$ is the scale height, $\alpha_{\rm in} > 0$ and $\alpha_{\rm out}<0$ are the asymptotic power law indices when $r \ll r_{\rm c}$ and $r \gg r_{\rm c}$, respectively. The critical radius can be converted to peak density radius, $r_{\textrm{max}}$, using Equation (1) of \citet{augereau99},
\begin{equation}\label{eq-rc-rmax}
r_{\textrm{max}}= \left(-\frac{\alpha_\textrm{in}+\beta}{\alpha_\textrm{out}+\beta} \right)^{(2\alpha_\textrm{in}-2\alpha_\textrm{out})^{-1}}r_\textrm{c}.
\end{equation}

The TWA~7 disk is nearly face-on, which makes constraining the vertical structure in Equation~\eqref{eq-powerlaw} less meaningful. We thus adopt $h=0.04$ from the vertical structure study for debris disks in \citet{thebault09}. The disk has an inclination of $\theta_{\rm inc}$, which is defined as the dihedral angle between the disk mid-plane and the sky. The position angle of the disk is $\theta_{\rm PA}$, which is defined as the position angle of the disk's semimajor axis---the one which is $90^\circ$ counterclockwise from the semiminor axis that is closer to Earth---measured from North to East.

\subsubsection{Scattering phase function}
Scattering phase function describes the intensity of light as a function of scattering angle, i.e., the angle measured from the incident light ray to the outgoing ray. With a ${\sim}10^\circ$ inclination for the TWA~7 disk \citep{olofsson18}, we are only able to study the scattering phase function of the system in a limited angle range (i.e., from ${\sim}80^\circ$ to ${\sim}100^\circ$). 

Motivated by the fact that scattering phase functions can be described with a single parametric function in such a limited angle range \citep[e.g.,][]{hedman15}, we adopt the parametric phase function in \citet{hg41} for the scatterers seen in total intensity in STIS and NICMOS,  
\begin{equation}\label{eq-hg}
I_{\rm tot}(\theta) = \frac{1-g^2}{4\pi (1+g^2-2g\cos\theta)^{3/2}},
\end{equation}
where $\theta$ is the scattering angle that is defined as the angle between the incident and emergent rays, $g\in[-1, 1]$ is a parameter that ranges from backward scattering when $g\leq0$ to forward scattering when $g\geq0$.  

For the SPHERE \Qphi\ image in polarized light, we have a Rayleigh-like polarization fraction term to modify the Henyey--Greenstein total intensity phase function \citep[e.g.,][]{engler17, olofsson18},
\begin{equation}\label{eq-ray-mod}
I_{\rm pol}(\theta) = I_{\rm tot}(\theta) \times \frac{1-\cos^2\theta}{1+\cos^2\theta}.
\end{equation}

\subsubsection{Approach}
The \citet{millarblanchaer15} code can satisfy the above requirements in geometric structure\footnote{We have adjusted the code using Equation~\eqref{eq-powerlaw}.} and scattering phase function, we thus use it to produce synthetic images for optically thin disks. To scale the overall brightness of an output disk image to match the observed data, we multiply the output by a parameter $f_\textrm{flux}$. To take advantage of parallel computation, we distribute the calculations on a computer cluster using the {\tt DebrisDiskFM} package \citep{ren19}, and explore the parameter space with a Monte Carlo Marko chain (MCMC) approach using the {\tt emcee} package \citep{emcee}. 

To obtain the best-fit disk parameters for the observations, we maximize the following log-likelihood function under independent Gaussian distribution,
\begin{align*}\label{eq-loglike}
\ln\mathcal{L}\left(\bm{\Theta}\mid X_{\rm obs}\right) = &-\frac{1}{2}\sum_{i=1}^{N}\left(\frac{X_{{\rm obs}, i} - X_{{\rm model}, i}}{\sigma_{{\rm obs}, i}}\right)^2\\
	&- \sum_{i=1}^{N}\ln\sigma_{{\rm obs}, i} - \frac{N}{2} \ln(2\pi), \numberthis
\end{align*}
where $\bm{\Theta}$ denotes the set of disk parameters (i.e., $\theta_\textrm{inc}$, $\theta_\textrm{PA}$, $\alpha_\textrm{in}$, $\alpha_\textrm{out}$, $r_\textrm{c}$, $g$), $X$ is a disk image with $N$ pixels that is either an actual observation ($X_{\rm obs}$) or a synthetic model ($X_{\rm model}$), $\sigma$ is the uncertainty map that has the same dimension of $X$. Aiming at only establishing the structure for the TWA~7 system in this paper, we do not adjust Equation~\eqref{eq-loglike} to address correlated noise using a covariance-based log-likelihood function as in \citet{wolff17}.

The STIS and SPHERE disk images are not expected to suffer from data reduction biases in cRDI or PDI, while the NICMOS image do in MRDI with NMF. We therefore adopt different modeling approaches to retrieve the disk parameters. For the SPHERE and STIS images, we maximize Equation~\eqref{eq-loglike} by directly modeling the disk images. For NICMOS, we use forward modeling: we subtract disk models from the original observations, and perform NMF reduction using the previously calculated NMF components in Section~\ref{sec-dr-nicmos} to minimize the residuals. Such a negative injection process for NICMOS effectively changes Equation~\eqref{eq-loglike} by substituting $X_{\rm obs}$ with the residual map after data reduction, and $X_{\rm model}$ with zero.

\subsection{Disk Modeling}\label{sec-disk-modeling}
\begin{figure}[htb!]
	\centerfloat
	\includegraphics[width=0.52\textwidth]{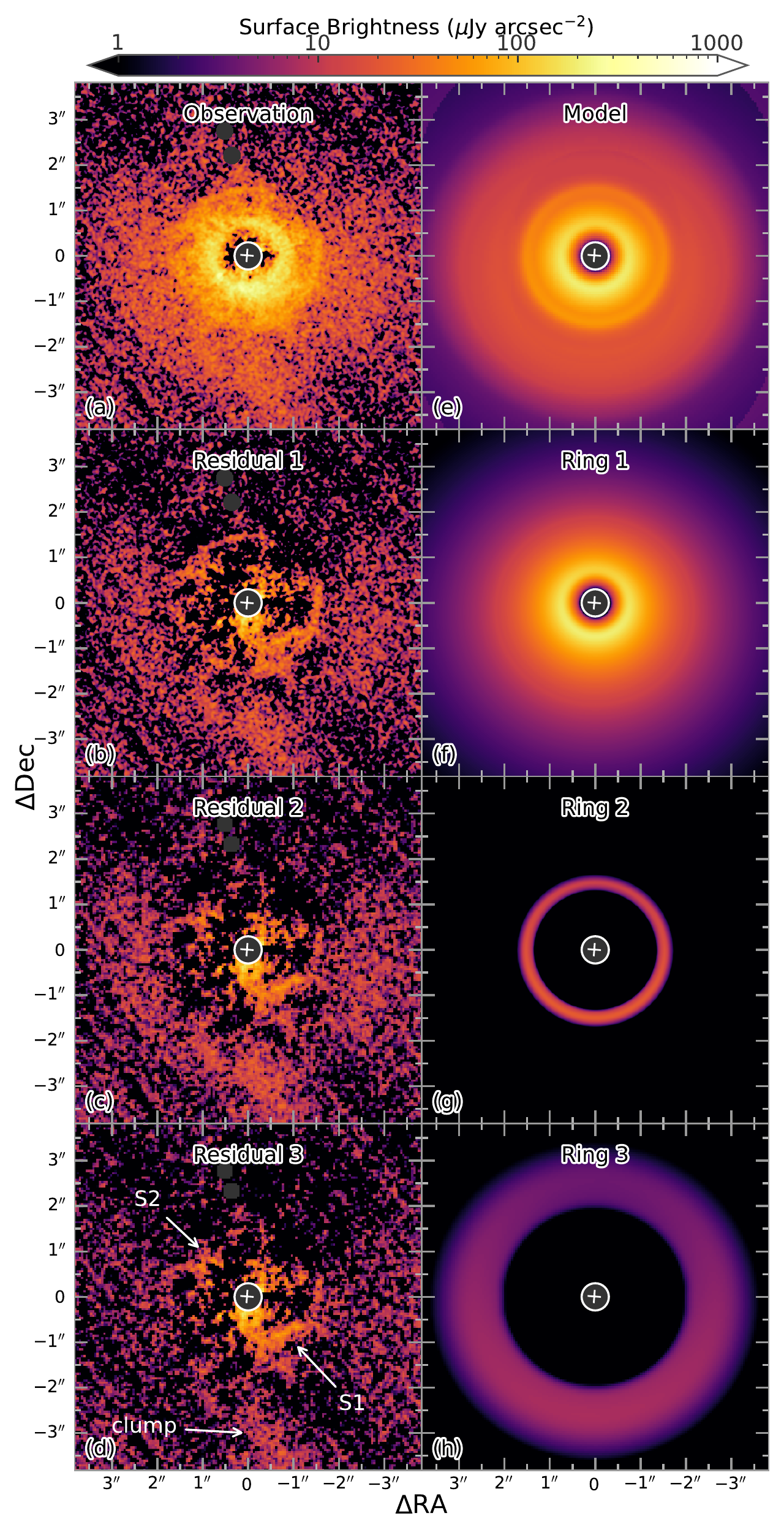}
    \caption{Modeling of the SPHERE observation in Section~\ref{sec-model-sphere}. (\textbf{a}) is the SPHERE \Qphi\ image and (\textbf{e}) is the best-fit model. The residual maps in (\textbf{b}), (\textbf{c}) and (\textbf{d}) have the individual ring models in (\textbf{f}), (\textbf{g}) and (\textbf{h}) cumulatively removed from (\textbf{a}). Notes: each ring has only one double-powerlaw component described by Equation~\eqref{eq-powerlaw}; see Figure~\ref{fig-structure-sphere} for the surface density radial profiles.
    }
    \label{fig2}

(The data used to create this figure are available in the ``anc'' folder

\hspace{-2.8in}on arXiv.)
\end{figure}

\begin{figure*}[htb!]
	\centerfloat
	\includegraphics[width=0.5\textwidth]{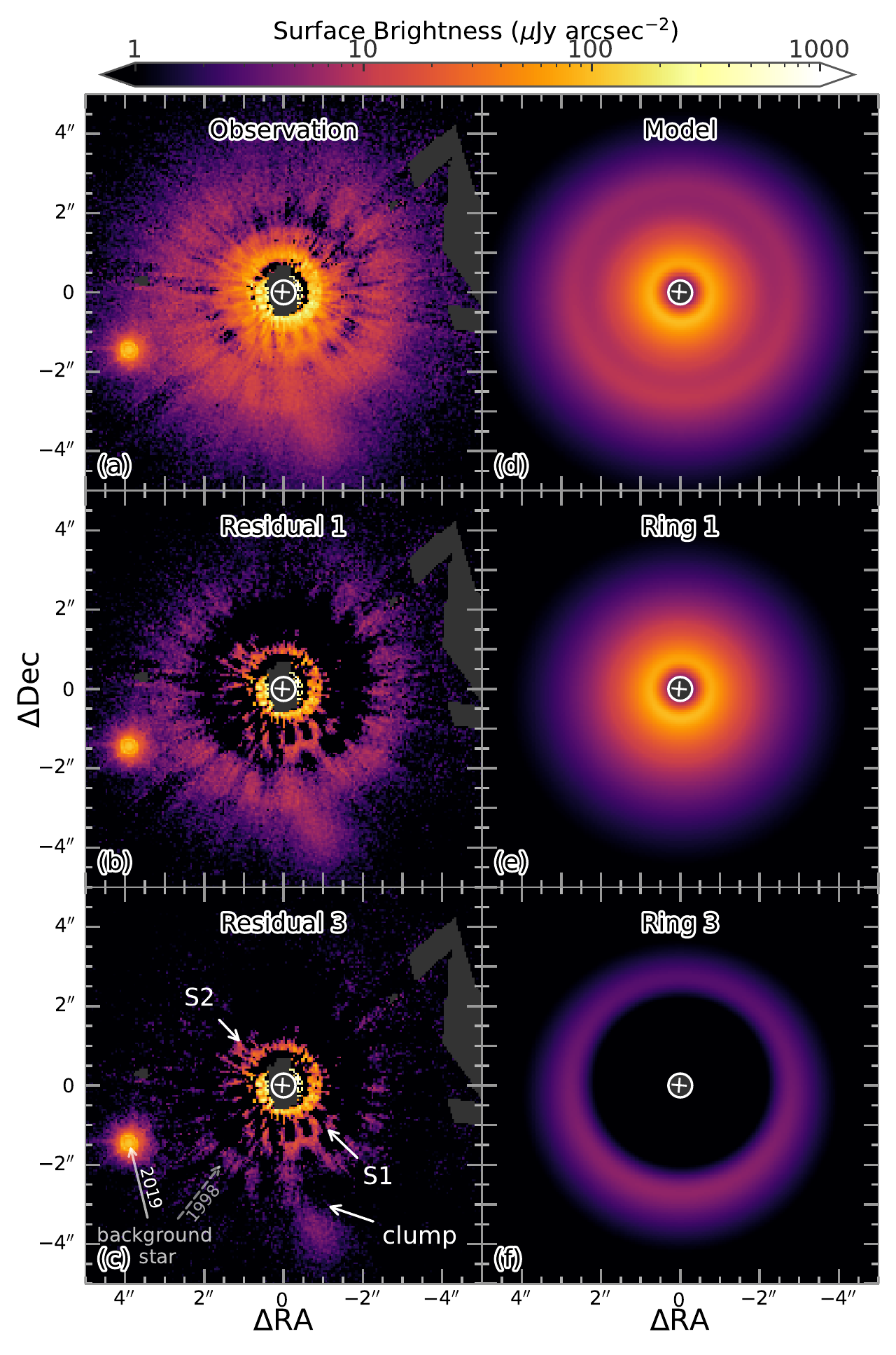}
    	\includegraphics[width=0.4625\textwidth]{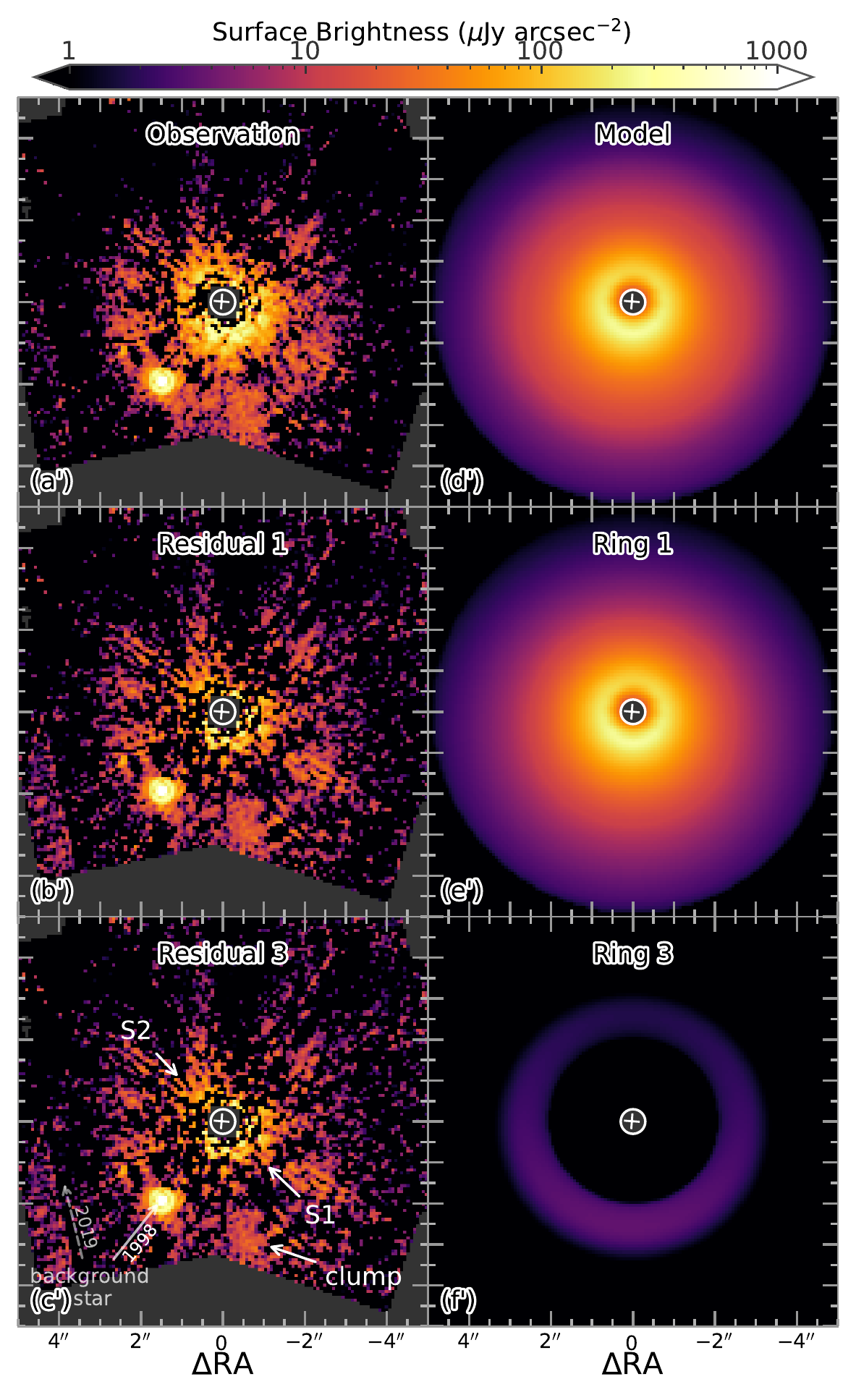}
    \caption{Modeling of the STIS and NICMOS observations in Sections~\ref{sec-model-stis} and \ref{sec-model-nicmos}, see (\textbf{a}) -- (\textbf{f}) and (\textbf{a'}) -- (\textbf{f'}), respectively. For STIS, (\textbf{a}) is the observation and (\textbf{d}) is the best-fit model; the residual maps in (\textbf{b}) and (\textbf{c}) have the ring models in (\textbf{e}) and (\textbf{f}) cumulatively removed from (\textbf{a}). For NICMOS, we have adopted a forward modeling strategy to fit the three rings simultaneously, see see (\textbf{d'}) for the best-fit model and (\textbf{c'}) for the corresponding residuals. For the purpose of illustration only, (\textbf{b'}) and (\textbf{e'}) are the best-fit Ring~1 residual and model. Note: there is only one double-powerlaw component in each ring model; Ring~2 fit was attempted for both instruments, but was not confidently recovered.
    }
    \label{fig3}
\hspace{-3.25in}(The data used to create this figure are available in the ``anc'' folder on arXiv.)
\end{figure*}

We perform disk modeling to establish the architecture of the TWA~7 system and compare the data across different instruments. To minimize the impact from residual physical structures on disk parameters (e.g., the spiral and secondary ring in \citealp{olofsson18}) for a ring, we exclude certain regions in calculating the likelihood function in Equation~\eqref{eq-loglike}. After iteratively performing disk modeling and inspecting the residuals with different ignored regions, we detect possibly physical structures with pixelwise average signal to noise ratio (S/N) $> 1$: the background star, two possible spiral arms, and a clump. We thus converge on masking out the S1 (the arm reported in \citealp{olofsson18}), S2 (a low S/N structure in STIS), clump, and background star regions in calculating the likelihood in disk modeling, see the annotations in Figures~\ref{fig2} and \ref{fig3}. 

To model the rings in the TWA~7 system, we have attempted modeling the system by simultaneously exploring all the disk parameters in Equation~\eqref{eq-loglike} for the three ring components in Figure~\ref{fig1}, however, combining multi-dimensional MCMC exploration with disk modeling is computationally prohibitive even with a computer cluster. We have also treated the rings as spatially separated components in different annuli from the star, however the rings could not be depicted by the double-powerlaw description in Equation~\eqref{eq-powerlaw}. We detail our adopted modeling procedure below, and present the $50$th$\pm34$th percentiles for the retrieved disk parameters in Table~\ref{tab-mcmc-params}.

\begin{deluxetable}{c|c|ccc}[thb!]
\tablecaption{$50$th$\pm34$th percentiles of TWA~7 ring parameters\label{tab-mcmc-params}}
\tablehead{ 
{Instrument}			&		{Parameter} 	& {Ring~1} 		& {Ring~2}\tablenotemark{e} 			&  {Ring~3}\tablenotemark{f} 	
 }
\startdata 
\multirow{7}{*}{SPHERE}	& $\theta_{\rm inc}$\tablenotemark{a} 	&$12\fdg91_{-0\fdg19}^{+0\fdg18}$	& $\cdots$ 				& $\cdots$ \\
					& $\theta_{\rm PA}$\tablenotemark{a} 	&$-94\fdg3_{-0\fdg5}^{+0\fdg5}$ 	& $\cdots$ 				& $\cdots$\\ \cline{2-5}
				 	& $\alpha_{\rm in}$					& $6.34_{-0.08}^{+0.07}$ 			& $13.4_{-0.6}^{+0.6}$ 		& $2.1_{-0.5}^{+0.6}$\\ 
					& $\alpha_{\rm out}$					& $-1.739_{-0.013}^{+0.013}$ 		& $-23.8_{-1.2}^{+1.4}$ 		& $-8.0_{-1.6}^{+1.2}$\\ 
					& $r_{\rm c}$ (au) 					& $24.49_{-0.04}^{+0.04}$ 		& $52.38_{-0.12}^{+0.12}$ 	& $101_{-3}^{+2}$\\
					& $r_{\rm max}$\tablenotemark{b} (au) 	& $28.12_{-0.02}^{+0.01}$ 		& $51.88_{-0.02}^{+0.02}$ 	& $93.1_{-0.4}^{+0.2}$\\
					& $g$\tablenotemark{c} 				& $0.395_{-0.009}^{+0.009}$ 		& $0.921_{-0.015}^{+0.013}$ 	& $0.94_{-0.04}^{+0.02}$\\  \hline
\multirow{5}{*}{STIS} 	& $\alpha_{\rm in}$& 				$6.98_{-0.14}^{+0.15}$ 			& $\cdots$ 				& $9.61_{-0.16}^{+0.15}$\\ 
					& $\alpha_{\rm out}$& 				$\cdots$\tablenotemark{d} 		& $\cdots$ 				& $-3.76_{-0.02}^{+0.02}$\\ 
					& $r_{\rm c}$ (au)  					& $\cdots$\tablenotemark{d} 		& $\cdots$ 				& $92.15_{-0.16}^{+0.16}$\\ 
					& $r_{\rm max}$\tablenotemark{b} (au) 	& $27.891_{-0.007}^{+0.008}$		& $\cdots$ 				& $96.56_{-0.05}^{+0.02}$\\ 
					& $g$							& $0.278_{-0.009}^{+0.009}$ 		& $\cdots$ 				& $0.959_{-0.006}^{+0.007}$\\  \hline
\multirow{4}{*}{NICMOS} 	& $\alpha_{\rm in}$\tablenotemark{d}		& $\cdots$  					& $\cdots$ 				& $\cdots$ \\ 
					& $\alpha_{\rm out}$\tablenotemark{d}	& $\cdots$ 					& $\cdots$ 				& $\cdots$ \\ 
					& $r_{\rm c}$\tablenotemark{d} (au)  		& $\cdots$			 		& $\cdots$ 				& $\cdots$ \\ 
					& $g$							& $0.87_{-0.02}^{+0.02}$  		& $\cdots$ 				& $0.96_{-0.11}^{+0.02}$ \\ 
\enddata
\tablecomments{
\tablenotetext{a}{Retrieved for Ring~1, and fixed for other instruments and rings.}
\tablenotetext{c}{Modified by Rayleigh scattering in Equation~\eqref{eq-ray-mod} for SPHERE data.}
\tablenotetext{b}{Peak surface density radius calculated using Equation~\eqref{eq-rc-rmax}.}
\tablenotetext{d}{Fixed to SPHERE best-fit values.}
\tablenotetext{e}{Ring~2 is not recovered in modeling STIS and NICMOS due to resolution and data quality.}
\tablenotetext{f}{Ring~3 parameters in SPHERE are retrieved from binned image (i.e., $4\times4$~pixel to $1$~bin).}
}
\end{deluxetable}

\subsubsection{SPHERE}\label{sec-model-sphere}

We start with modeling the SPHERE image that best resolves the first two rings in \citet{olofsson18}. Given that the SPHERE observation reveals two radial components that have similar intensity in the star-illumination--corrected radial profile of \citet{olofsson18}, we model the rings sequentially: we perform disk modeling for one ring, then subtract its best-fit model from the data and fit for another ring. To obtain the best-fit parameters for the system, we convolve the models with a two dimensional Gaussian that has a standard deviation of 2 pixels, and compare the convolved models with the convolved \Qphi\ image.

First, we model the inner ring at ${\sim}0\farcs8$ (i.e., ``Ring~1'') in \citet{olofsson18} within a $184\times184$ pixel region centered at the star. To minimize the impact from the residual light that is close to the coronagraph, we mask out a circular region with a radius of $24$ pixel centered at the star ($3$ times the size of the IWA). In addition, we ignore the region that hosts non-disk signals in calculating the log-likelihood \citep[e.g.,][]{wang20}, thus reducing the potential impact from the S1 spiral arm in \citet{olofsson18} on the retrieved disk parameters.

Second, we remove the best-fit model for the inner ring from the SPHERE image, then model the secondary ring at ${\sim}1\farcs5$ (i.e., ``Ring~2'') within an annulus that is between 110 and 135 pixel ($1\farcs348$ to $1\farcs658$). We assume Ring~2 has identical inclination and position angle as Ring~1, and thus only explore the critical radius and radial power law indices in Equation~\eqref{eq-powerlaw}, and the Henyey--Greenstein $g$ parameter.

Third, we model the tertiary ring (i.e., ``Ring~3'') after removing the models for the first two rings, assuming identical inclination and position angle as Ring~1 and exploring identical parameters as for Ring~2. We bin the observation data in a $4{\times}4$ pixel region to $1$ bin to increase calculation efficiency and S/N. In our first attempt, we could recover a southern clump that is evident in the STIS data, we thus mask out that region in the modeling to reduce its impact on the retrieved disk parameters in disk modeling. 

We present in Figure~\ref{fig2} our sequential modeling results for the three rings in the SPHERE observation, see Section~\ref{sec-3rings} for the corresponding radial profiles. In comparison with the retrieved disk parameters in \citet{olofsson18}, our best-fit values in Table~\ref{tab-mcmc-params} are within their $1\sigma$ credible intervals. Nevertheless, we have masked out S1 and focused on the region interior to Ring~2  (spatial extent determined by minimizing the residuals around Ring 2) in our fitting for Ring~1, and thus our best-fit values deviate from \citet{olofsson18}. On the one hand, by ignoring S1, the best-fit position angle of the major axis reported here is shifted $5^\circ$ clockwise\footnote{Note the difference in the definition, $\theta_\textrm{PA}$ here is $180^\circ-\phi$ in \citet{olofsson18}.}. This can be explained by the fact that S1 is located at the southwest region, and the S1 tip have contributed to the fitting of Ring~1 in \citet{olofsson18}. On the other hand, by additionally ignoring Ring~2, the best-fit power law index for the tail of Ring~1 is $-1.7$ here, which is steeper than the \citet{olofsson18} best-fit of $-1.5$; this has been expected then. We do not discuss other parameters since they are less impacted by S1 and Ring~2. What is more, we do not compare the credible intervals since we have not adopted the kernel density approach, where the kernel widths depend on user inputs, as in \citet{olofsson18}.

\subsubsection{STIS}\label{sec-model-stis}
We follow the above SPHERE procedure of sequentially modeling the ring components to model the system architecture in STIS. To simulate the responses of the STIS detector, we convolve the models with a  {\tt TinyTim} PSF \citep{tinytim}\footnote{\url{http://tinytim.stsci.edu}} that has an effective temperature of $4018$~K for TWA~7 \citep{gaiadr2}. See Figure~\ref{fig3} for the detailed sequential modeling results.

After attempting to vary $\alpha_{\rm out}$ for Ring~1, we do not obtain statistically different result from the SPHERE value, we thus fix it to be the best-fit value from SPHERE for STIS. Despite Ring~2 being visible by eye in Figure~\ref{fig1}c, we are limited by residual noises in disk modeling, and thus we cannot recover Ring~2 in STIS modeling beyond $1\sigma$ in $f_\textrm{flux}$. We therefore do not report the disk parameters for Ring 2. 

In comparison with the SPHERE residuals in Figure~\ref{fig2}d, we tentatively identify in STIS an additional spiral, S2, with an average pixelwise S/N of $2.3$. If S2 does not originate from residual speckles in Figure~\ref{fig3}c, then STIS is able to better resolve it than SPHERE. Such a scenario is likely, since that the exposure time with STIS is $11$ times that with SPHERE, and that STIS observes the system in total intensity while SPHERE in polarized light.

\subsubsection{NICMOS}\label{sec-model-nicmos}
NICMOS and SPHERE cover nearly identical wavelengths, we thus use the SPHERE spatial distribution to model the NICMOS image. To address the difference between total intensity and polarized phase functions in Equations~\eqref{eq-hg} and \eqref{eq-ray-mod}, we only vary the $g$ parameters and brightness levels for the three NICMOS rings. We rotate a disk model according to the telescope roll angles of the NICMOS observations, then convolve the rotated disk models with a {\tt TinyTim} PSF for the F160W filter. In forward modeling, we subtract them from the NICMOS observations, and reduce the disk-subtracted observations with previously generated NMF components.

We simultaneously model the three rings in NICMOS to minimize potential overfitting biases in forward modeling, see the best-fit results in Figure~\ref{fig3}. Limited by residual noises, we cannot recover Ring~2 in NICMOS beyond $1\sigma$ in $f_\textrm{flux}$. For the purpose of illustration only, we adopt the best-fit parameters for Ring~1, perform NMF forward modeling, and present the corresponding results in Figure~\ref{fig3}b'. In comparison with the retrieved disk parameters in \citet{choquet16}, where the disk parameters are loosely constrained using only NICMOS data, we are able to better constrain the $g$ parameter for Ring~1 using  the spatial distribution inferred from the SPHERE data. Nevertheless, given that the ${\sim}13^\circ$ inclination of this system only allows the study of the scattering phase function from $77^\circ$ to $103^\circ$, we note that a discussion of the Henyey--Greenstein $g$ parameter, given the large NICMOS pixel size and the contaminations on the rings, would be less informative, see, e.g., Section~\ref{sec-ring1-rayleigh-or-not}.

\section{Spatial Distribution}\label{sec-spatial-comp}
Using static models, our disk modeling reveals a layered architecture for the TWA~7 debris disk system.  The extended ring, Ring~1, whose density peaks at $28$~au dominates the overall surface density distribution from ${\sim}10$~au to ${\sim}200$~au. The secondary and tertiary rings, Ring~2 and Ring~3, peak at $52$~au and $93$~au, are superimposed onto the outskirts of Ring~1, see Section~\ref{sec-3rings}. In comparison with the locations of the TWA~7 rings from SED analysis \citep[e.g.][]{RiviereMarichalar13, bayo19}, we find that blackbody or Mie assumption cannot predict the rings in scattered light images with high accuracy \citep[see also][]{esposito20}, and correction factors are needed to account for such discrepancies \citep[e.g., sub-mm to blackbody:][]{pawellek21}.

There likely exist two spirals, S1 and S2 that are superimposed on Ring~1 and Ring~2, see Section~\ref{sec-spiral-motion} for the smoothed residuals. S1 has been tentatively detected in \citet{olofsson18}, and we now well-resolve it in both SPHERE and NICMOS data after disk modeling. S2 is tentatively resolved in STIS residuals, and it has an average pixelwise S/N of $2.3$; yet it is less resolved in SPHERE and NICMOS. 

Superimposed onto Ring~3, there exists a clump in all three observations, see Section~\ref{sec-clump}. We resolve this elliptical clump in all three instruments, and our STIS data best resolves the clump with with an area of ${\approx}150$~au$^2$. By establishing a $21$~yr span in this paper, this dusty clump is a component on the TWA~7 debris disk, since a background galaxy would have otherwise moved with respect to TWA~7. In comparison, this clump has been marginally resolved at ${\sim}2\sigma$ by ALMA in \citet{bayo19}.

\subsection{Overall structure}\label{sec-overall-struct}

\subsubsection{Three rings}\label{sec-3rings}

\begin{figure}[hbt!]
	\centerfloat
	\includegraphics[width=0.5\textwidth]{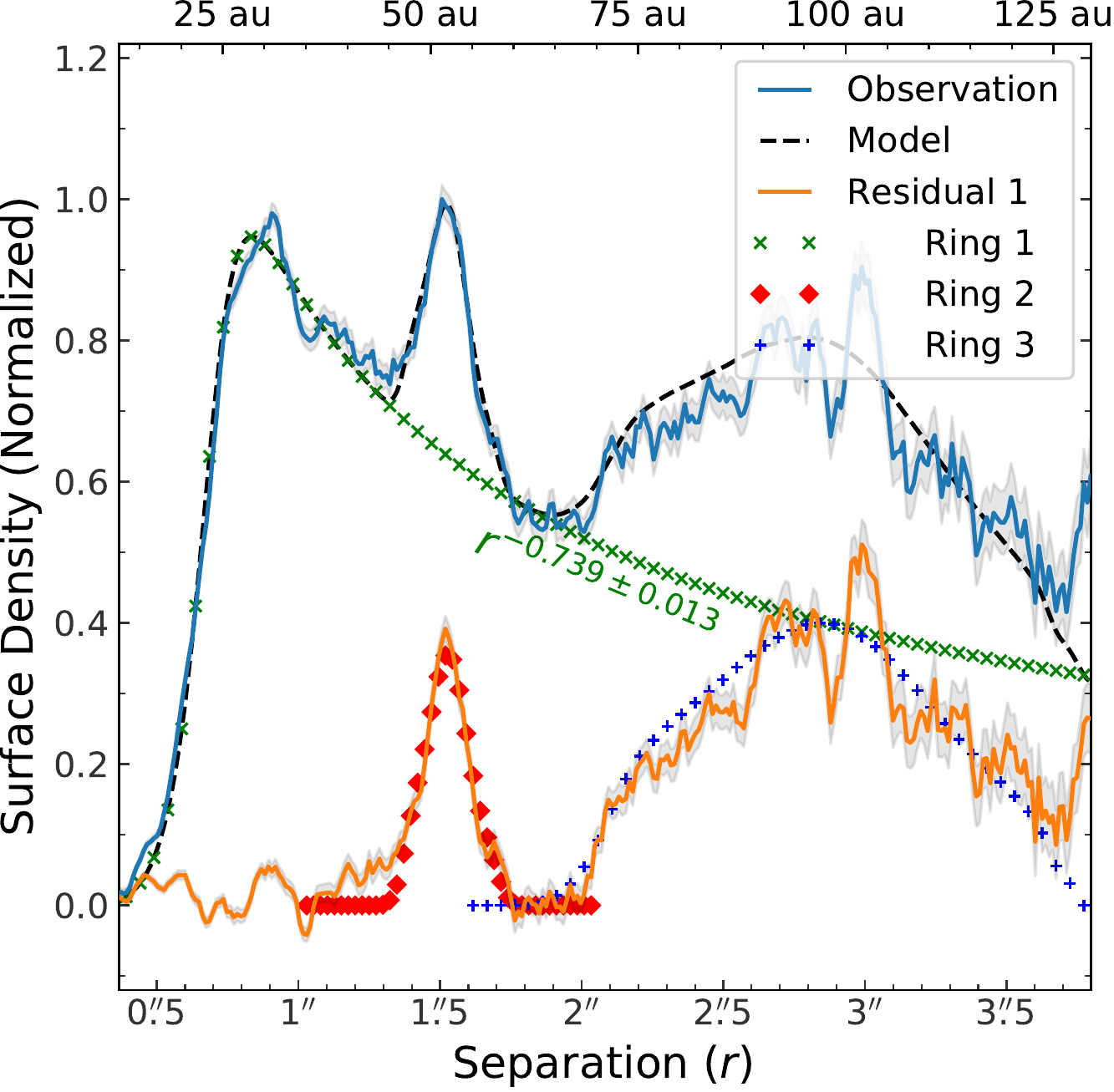}
    \caption{Surface density radial profiles, calculated by multiplying surface brightness radial profiles with squared stellocentric distances, for SPHERE data and models. Ring~1 dominates the surface density distribution of the system. Ring~2 and Ring~3 are of similar surface density: when they are superimposed onto the tail of Ring~1, the three ring components have similar levels of total surface density. See Figure~\ref{fig2} for the corresponding surface brightness distribution, and Table~\ref{tab-mcmc-params} for the parameter values in Equation~\eqref{eq-powerlaw}. Note: the surface density radial profiles are normalized by dividing the peak value from the observation.}
    \label{fig-structure-sphere}
\end{figure}

Our modeling shows that there are three rings in the TWA~7 system, with an extended one centered at ${\sim}0\farcs8$ dominating the overall surface brightness distribution as well as the surface density distribution. For the SPHERE data, we multiply the surface brightness for each location with its corresponding squared stellocentric distance to compare the surface density radial profile, see Figure~\ref{fig-structure-sphere}. 

Although Ring~1 dominates the surface brightness distribution of the system, we observe in Figure~\ref{fig-structure-sphere} that Ring~2 and Ring~3 have similar surface density in the radial profile for ``Residual 1'', which corresponds to the residual map after removing Ring~1 best-fit model from the SPHERE \Qphi\ image. When Ring~2 and Ring~3 are superimposed onto the tail of Ring~1, the three rings have similar levels of total surface density that peak sequentially at $0\farcs8$, $1\farcs5$, and $2\farcs8$.

We confirm the existence of Ring~1 and Ring~3 in NICMOS and STIS observations with disk modeling, see Figure~\ref{fig3}. In comparison, although we are able to resolve Ring~2 by eye in STIS in Figure~\ref{fig1}c, the radially extended spike structures between Ring~1 and Ring~2 may have led to overfitting for Ring~1 in STIS. The overfitting then overwhelms the Ring~2 signal in disk modeling. In addition, we are not able to resolve Ring~2 either by eye or from forward modeling in NICMOS. However, we observe a plateau at $1\farcs5$ in Figure~\ref{fig1}d that is indicative for the existence of Ring~2, yet this plateau may originate from joint effects from both the spirals and Ring~2.

\subsubsection{Ring 1 tail}\label{sec-ring1-tail}
The power law index for the tail of Ring~1 is $\alpha_\textrm{out} = -1.739_{-0.013}^{+0.013}$ from SPHERE disk modeling. The index does not have a significant difference between SPHERE and STIS observations. This suggests that the probed dust particles are likely well-mixed for Ring~1 tail in STIS and SPHERE wavelengths. In comparison, Equation (5) of \citet{thebault08} predicts the dominant particle size as a function of radial separation from the star. Nevertheless, the similar power law indices between SPHERE and STIS here suggest that the larger particles, which are probed by SPHERE and are the dominant sizes for closer-in regions, will continue dominate further out regions.

Using the relationship between $\alpha_\textrm{out}$ and surface density power law index $\Gamma_\textrm{out}$ in \citet{augereau99}, i.e., $\Gamma_\textrm{out}=\alpha_\textrm{out}+\beta$ where $\beta=1$ as in Equation~\eqref{eq-powerlaw}, the surface density power law index of the Ring 1 tail is $\Gamma_\textrm{out}=-0.739_{-0.013}^{+0.013}$, see the annotation on Figure~\ref{fig-structure-sphere}. This surface density is steeper than the \citet{olofsson18} index of $-0.52$, since we have our excluded Ring~2 in our fitting. Meanwhile, \citet{schneider18} and \citet{ren19} have observed similar indices of approximately $-0.7$ for the tails of the debris disks orbiting A0 star HR~4796~A and F5 star HD~191089. 

We discuss the meaning of the power law index of Ring 1 tail below assuming Ring 1 is the birth ring of small particles in the system. However, we caution that with a single birth ring, stellar winds around M stars can create multiple rings in shorter wavelengths, and the location of the brightest ring deviates from that of the birth ring \citep[e.g., Figure 5 of][]{pawellek19}.

The classical expectation for the surface density power law index of collision-dominated debris disk tails is $-1.5$ \citep[e.g.,][]{strubbe06, thebault08}, while the index is $-2.5$ when a disk is dominated by corpuscular and Poynting--Robertson drag from strong stellar winds \citep[e.g.,][]{strubbe06}. In  Section 4.3 of \citet{strubbe06}, the former is expected for $1 \lesssim \dot{M}_\textrm{star}/\dot{M}_\Sun \lesssim 10$, while the latter for $10^2 \lesssim \dot{M}_\textrm{star}/\dot{M}_\Sun \lesssim10^3$. The measured surface density power law index of $-0.7$ for M star TWA~7 here, however, deviates from the expectations, which makes the involvement of additional mechanisms necessary (e.g., companion radiation from HR~4796~B in HR~4796~A tail, or possible interstellar medium slowdown in HD~191089 tail). 

In comparison with M star AU~Mic, whose surface density power law index of $-1.5$ is explained in \citet{strubbe06} with $\dot{M}_\textrm{star}/\dot{M}_\Sun \lesssim 10$, the existence of Ring~2 and Ring~3 makes it a more complicated scenario for the surroundings of M star TWA~7. Nevertheless, the detected CO gas that extends to ${\sim}100$~au, or ${\sim}3\arcsec$, around TWA~7 in \citet{matra19} is likely a non-negligible source: gas can slow down the radially outward motion of the smallest dust particles; under this scenario, the slow-down explanation in \citet{ren19} might help interpret the power law index for the tail of Ring~1.

\subsubsection{Ring 3 peak}
The peak density radius, $r_\textrm{max}$, for Ring~3 is likely located outwards in the STIS data ($r_\textrm{max}=97$~au) than in the SPHERE data ($r_\textrm{max}=93$~au). In addition, the $\alpha_\textrm{out}$ power law index for Ring~3 is shallower in STIS ($\alpha_\textrm{out}=-3.76$) than in SPHERE ($\alpha_\textrm{out}=-8.0$). Both suggest that the smaller dust particles observed by STIS at ${\sim}0.6~\micron$ are likely located more extended than the larger dust particles observed by SPHERE at ${\sim}1.6~\micron$. 

The extension of small particles in STIS can be explained by the force balance between stellar radiation pressure and gravitational pull. For a particle with size $a$, the radiation force $F_\textrm{radiation} \propto a^2$, while the gravitational pull $F_\textrm{gravity}\propto a^3$. The ratio between the two forces is $F_\textrm{radiation} / F_\textrm{gravity} \propto a^{-1}$. Therefore, in comparison with the relatively large particles seen in SPHERE, the smaller particles observed by STIS likely have relatively (rather than absolutely) higher ratio between the radiation force and the gravitational pull. This relatively higher ratio helps small particles reach high-eccentricity orbits and thus arrive at further distances from the central star. Indeed, this is supported by the theoretical expectation that the optical depth is dominated by smaller particles at farther stellocentric separations, see, e.g., Equation~(5) of \citet{thebault08}.

\begin{figure*}[htb!]
	\centerfloat
	\includegraphics[width=\textwidth]{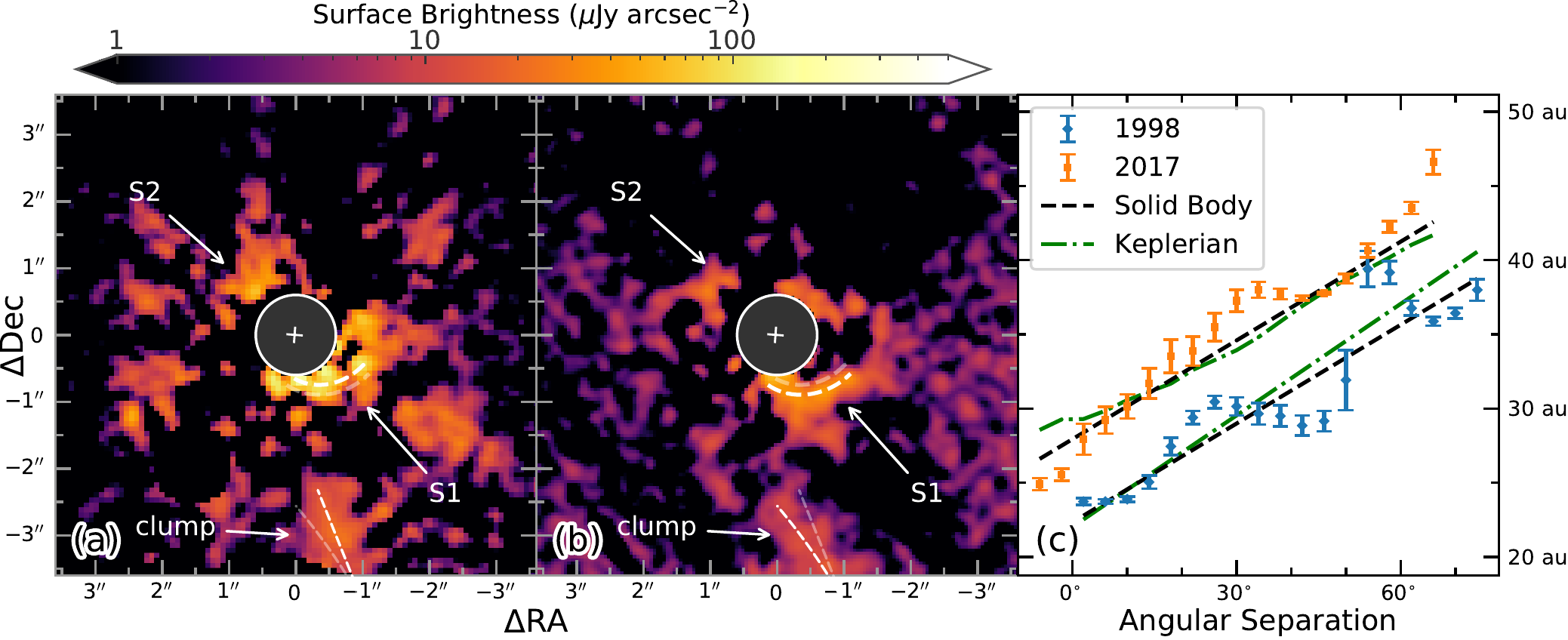}
    \caption{Spiral motion in Section~\ref{sec-spiral-motion}: smoothed residuals after disk modeling in (\textbf{a}) 1998 NICMOS, and (\textbf{b}) 2017 SPHERE. For S1, the dashed lines are the best-fits from solid body motion; for the clump, approximate centers: the white ones are for the corresponding observation, and the transparent ones are for the other. Panel (\textbf{c}) shows the location measurement and motion fitting for S1, and the clockwise motion can be explained by two spiral motion mechanisms. Notes: the spiral locations are measured from surface density maps, the error bars here are $3\sigma$, and the angular separation is calculated counterclockwise from the southeast semi-minor axis of the disk on the disk plane.
    }
    \label{fig-spiral-image}
\end{figure*}

\subsection{Possible moving components}\label{sec-addon}
\subsubsection{Spirals}\label{sec-spiral-motion}

We confidently resolve S1 in SPHERE after removing the disk models, and in NICMOS after smoothing the residuals after disk model removal, see Figure~\ref{fig-spiral-image}. We tentatively detect S2 in STIS with an average pixelwise S/N of $2.3$\footnote{In comparison, Section 3.6 of \citet{debes19} uses average S/N of 1 to calculate the detection limit of extended structures (specifically, rings) for STIS.} after removing the disk model, and marginally observe it in SPHERE after disk removal, see Figure~\ref{fig-spiral-image}b. For S2, our STIS model might be influenced by quasi-static noise near the coronagraph. In addition, S2 is located in the northern region of the TWA~7 system, which is at the far side from Earth in our disk modeling: the marginal detection of S2 in SPHERE and NICMOS is thus likely caused by scattering effects and less observation time. Specifically, in comparison with forward scattering, less light is scattered in backward scattering. Therefore, we only quantify the S1 motion between the 1998 NICMOS and 2017 SPHERE observations. 

To measure the motion of the S1 spiral, we follow the procedure in \citet{ren20b}. Specifically, in the surface density distribution map on the disk plane, we first fit Gaussian profiles to the radial profile at each angular position to obtain the peak position for the arm, then use dummy variables as proxies to obtain morphological parameters under polynomial description for the arm, and simultaneously quantify the arm motion rates under two hypotheses (i.e., either planet-driven motion or local Keplerian motion; \citealp{ren20b}). In the planet-driven scenario, the entire spiral moves as a solid body, and the spiral pattern motion traces the orbital motion of the driver; in the local Keplerian motion scenario, the motion of the spiral is faster when its location is closer to the star. To minimize systematics in pixel size, we have convolved the residual images with a two dimensional Gaussian that has a standard deviation of $75.65$~mas (i.e., 1 NICMOS pixel) for the images in Figure~\ref{fig-spiral-image}. We analyze the deprojected surface density maps on the disk plane to minimize stellar illumination and projection effects. We fit S1 in polar coordinates with $p$-degree polynomials as in \citet{ren20b}, and obtain minimized Akaike and Schwarz information criteria (AIC and SIC) at $p=1$, which corresponds to linear description of an arm in polar coordinates in the solid body motion scenario.

\begin{deluxetable}{c|c|c}[thb!]
 \setlength{\tabcolsep}{2pt}
\tablecaption{Pattern motion measurement for S1 \label{tab2}}
\tablehead{ 
Motion Pattern  &  Parameter & Value
 }
\startdata 
\multirow{5}{*}{Solid Body} & Rotation Rate (yr$^{-1}$) &$1\fdg3\pm0\fdg6$ \\
            & Driver Location\tablenotemark{a} (au) & $33_{-10}^{+19}$ \\
            & Orbital Period\tablenotemark{a} (yr) & $280_{-90}^{+240}$\\ \cline{2-3}
            & AIC & 5.68 \\
            & SIC & 8.95 \\ \hline
\multirow{4}{*}{Local Keplerian} & Rotation Rate (yr$^{-1}$) & $(1\fdg0\pm0\fdg6)\times(1\arcsec/r)^{3/2}$\tablenotemark{c} \\
            & Enclosed Mass\tablenotemark{b} ($M_\sun$) & $0.3^{+0.5}_{-0.3}$ \\ \cline{2-3}
            & AIC & 5.98 \\
            & SIC & 9.16 \\
            \enddata
\tablecomments{\tablenotetext{a}{The driver has a circular orbit along the midplane of the disk around a $0.46_{-0.10}^{+0.07}~M_\sun$ central star.}
\tablenotetext{b}{Enclosed mass within $20$~au (i.e., star and disk mass combined), as inferred from local Keplerian motion.}
\tablenotetext{c}{$r$ is the stellocentric separation in units of arcsec.}
}
\end{deluxetable}

We present the motion rates of S1 under the two mechanisms in Table~\ref{tab2}. To account for possible instrumental and data reduction uncertainties (e.g., NICMOS centering, focal plane mask location, NMF overfitting, surface brightness distribution difference in total intensity and polarized light: \citealp{ren18b}), we have adopted a relatively large $12^\circ$ uncertainty for NICMOS (i.e., ${\sim}3$ times the size of a NICMOS pixel at $1''$) when propogating the errors.

The measured motion direction of the S1 spiral is clockwise, which is consistent with the rotation of the CO gas in \citet{matra19} where the east side of the disk moves away from Earth. To compare the two motion mechanisms for S1, we calculate the AIC and SIC for both mechanisms. Both criteria are modifications to the classical $\chi^2$ statistic by adding penalty terms to dissuade excessive use of free parameters and avoid overfitting, and the model with the smallest AIC or SIC value is adopted as the best-fit model. Although the planet-driven model is slightly preferred, the difference in both information criteria is less than 1.\footnote{Under independent Gaussian noise assumption. The difference is smaller when the noises are correlated.} In comparison with the classical threshold of $10$ for model selection \citep[e.g.,][]{kass95}, we cannot distinguish the two motion mechanisms here. 

To investigate the nature of S1, we recommend follow-up SPHERE observations that probe the system with the same instrument and identical setup. When re-observed after 2021, a ${>}4$ year timeline will be established to help determine the motion mechanism for S1. Although it would be a shorter timeline than that presented in this paper, the identical setup and instrument can better constrain spiral motion: \citet{xie21} has recently demonstrated that PDI data with 1 year separation using SPHERE is sufficient in constraining the motion mechanism for the spiral arms in the protoplanetary disk surrounding SAO~206462. Nevertheless, we note that other mechanisms can excite spirals in debris disks \citep[e.g.,][]{sefilian21}.

\subsubsection{Ring~2}\label{sec-ring2}
We investigate the Ring~2 observations in the three instruments to explore its formation and motion mechanism. On the one hand, given that SPHERE best resolves Ring~2, we can inspect its surface brightness and thus optical depth distribution. On the other hand, given that we have visually detected Ring~2 in both SPHERE and STIS images, and that the radial profile for NICMOS reveals the possible existence of Ring~2, we can study its motion using the observations in all instruments that span from 1998 to 2019. Here we present two possible simple scenarios that can be tested using motion measurements with another SPHERE observation.

\begin{figure*}[hbt!]
	\centerfloat
	\includegraphics[width=\textwidth]{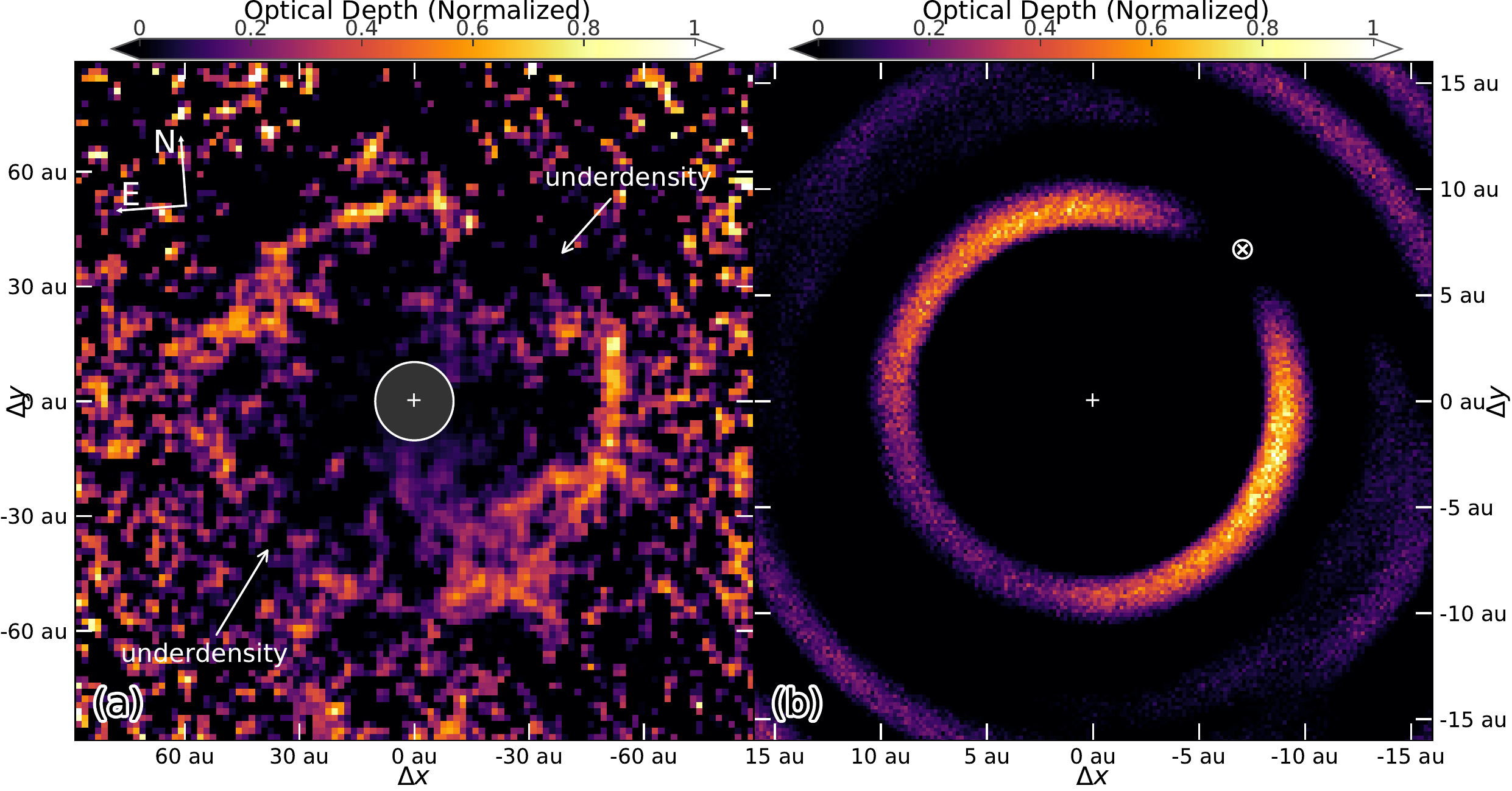}
    \caption{Mean-motion resonance scenario for Ring 2 in Section~\ref{sec-ring2}. (\textbf{a}) Deprojected optical depth of SPHERE residuals after removing the best-fit models for Ring~1 and Ring~3. The residuals resemble resonance structures induced by a planetary perturber (e.g., Figure~2b of \citealp{ozernoy00}). (\textbf{b}) High-pass-filtered and Poisson-noise-added simulation of mean motion resonance for collisionless dust particles whose radiation-to-gravity force ratio is $0.0023$ in a system that has a $5~M_\textrm{Earth}$ planet, marked by $\otimes$, located at $10$~au from the Sun  \citep{stark08}.}
    \label{fig-ring2-residuals}
\end{figure*}

\paragraph{Static image:  a resonance structure?}
We remove the best-fit models for Ring~1 and Ring~3 from SPHERE, correct for not only the inverse-square law of stellar illumination but also the best-fit scattering phase function for Ring~2 to obtain the optical depth for the residuals \citep[e.g., the procedure in][except we here obtain the Henyey--Greenstein $g$ parameter from disk modeling]{stark14}. We then deproject the disk to a face-on view, rotate the deprojected optical depth map to  align the major axis with $x$-axis, and present the optical depth map that is normalized at the peak in Figure~\ref{fig-ring2-residuals}. 

\begin{figure*}[hbt!]
	\centerfloat
	\includegraphics[width=\textwidth]{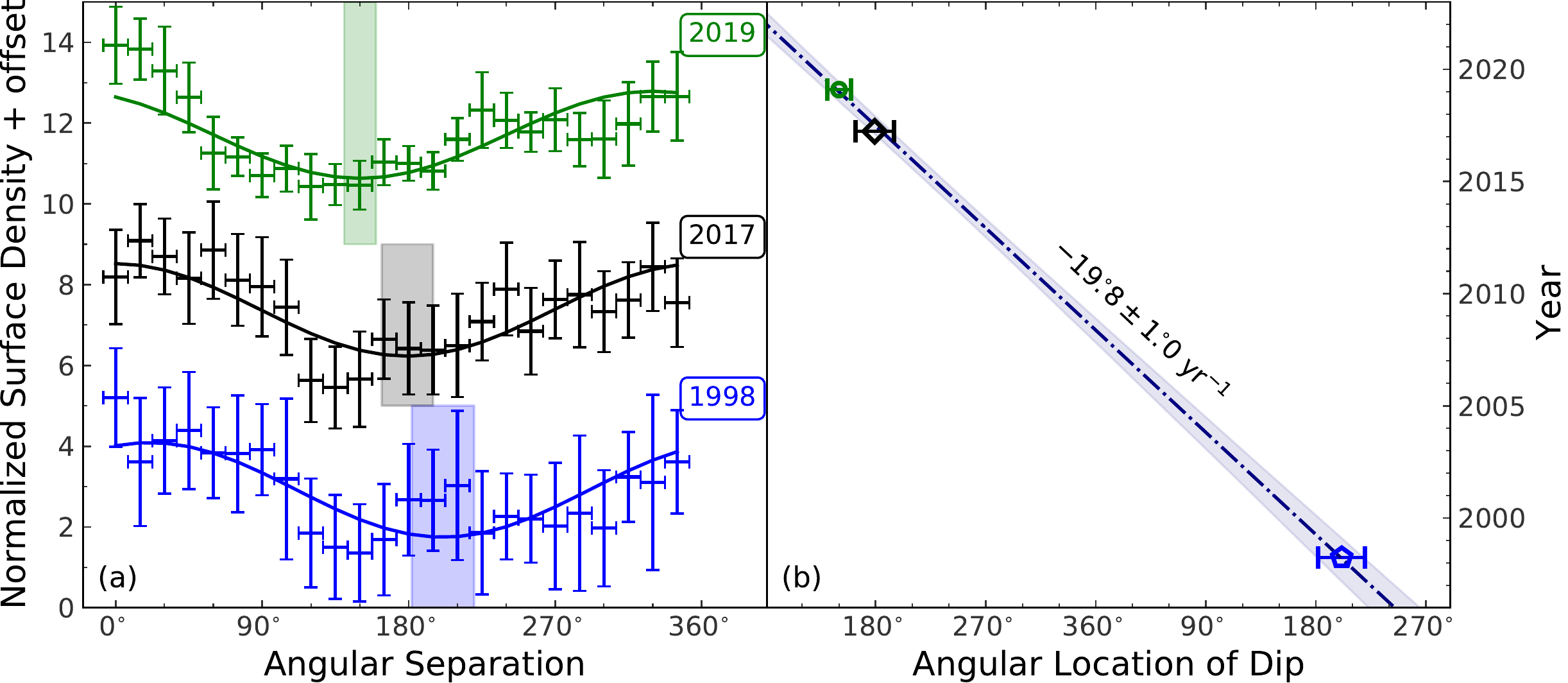}
    \caption{Shadowing effects scenario for Ring 2 in Section~\ref{sec-ring2}. (\textbf{a}) Azimuthal profiles for the deprojected surface density maps at the location of Ring~2, and cosine fits to them. The best-fit $\pm 1\sigma$ locations for the dip are shaded. (\textbf{b}) The motion of the dip across the three observations can be explained with a shadow that moves at $19\fdg8\pm1\fdg0$~yr$^{-1}$ clockwise. Under this scenario, a shadowing-casting disk under Keplerian motion at $5.3_{-0.5}^{+0.4}$~au can explain the apparent motion for the underdensity region of Ring~2. Note: the 1998 data are prone to reduction artifacts, and the motion here could completely change with follow-up observations.}
    \label{fig-ring2-dip-motion}
\end{figure*}

The relative under-density of materials in the northwest and southeast regions resembles resonance structures induced by exoplanets on debris disks \citep[e.g.,][]{ozernoy00, stark08, gozdziewski20}. The overall geometry qualitatively resembles Figure~2b of \citet{ozernoy00}, where the $2{:}1$ and $3{:}2$ resonances are in equal proportions for a system whose planet-to-star mass ratio is $0.25~M_{\textrm{Jupiter}}/M_\sun$. Were TWA~7 system with a similar planet-to-star mass ratio, the planet should be located in the northwest underdensity region at $r\approx50$~au, and its mass is ${\sim}2~M_{\textrm{Neptune}}$. We calculate in Section~\ref{sec-contrast} the contrast limits for the NICMOS observations, and such a planet is beyound detection using the NICMOS data.

For a qualitative illustration of the possible resonance structures, we inspect the Exozodi Simulation Catalog\footnote{\url{https://asd.gsfc.nasa.gov/Christopher.Stark/catalog.php}} and obtain the distribution of collisionless particles that are shepherded by a $5~M_\textrm{Earth}$ planet located at $10$~au from the Sun in \citet{stark08}. Noticing that we have removed in Figure~\ref{fig-ring2-residuals}a the Ring~1 model which controls the overall surface density for the TWA~7 system, we first smooth the simulation with a Gaussian kernel that has a standard deviation of $0.7$~au, then remove the extended structures in the simulation by subtracting from it a Gaussian-smoothed version of the simulation (standard deviation: $1.4$~au, i.e., high-pass filtering the data), see the resulting Poisson-noise-added image in Figure~\ref{fig-ring2-residuals}b. Under this scenario, although we have spatially separated Ring~2 and Ring~3 in our modeling, the two rings can be physically produced simultaneously by the same perturbing planet (see, e.g., Figure 6 of \citealp{stark08}). 

Despite difficulties in directly imaging a planetary perturber for TWA~7, follow-up SPHERE observations will help investigate the existence of the underdensity regions, as well as their motion. Specifically, the motion of the underdensity regions in Figure~\ref{fig-ring2-residuals}a should trace the orbital motion of the planet. Nevertheless, we caution that our resonance inference here is solely based on a static image of Ring 2 and we have ignored Ring 1 in our qualitative comparison: a more dynamically motivated approach is needed to study all the rings and explore the resonance structure for the system.

\paragraph{Multi-epoch images: a shadowing structure?}
We calculate the azimuthal profiles for Ring~2 to explore the apparent azimuthal motion of the ring \citep[rather than onsite physical motion, see, e.g.,][]{debes17}. Although we do not recover Ring~2 in \textit{HST} disk modeling, we do observe plateaus in the corresponding radial profiles. For a stellocentric separation between $46$~au and $59$~au, in which Ring~2 resides, we calculate the median and standard deviation for the deprojected surface density maps with an angular step of $15^\circ$. To measure the motion for the underdensity region, we first fit cosine profiles to the azimuthal profiles \citep[e.g.,][]{debes17}, then calculate linear angular motion for the dips, see Figure~\ref{fig-ring2-dip-motion}.

The 2017 SPHERE and 2019 STIS data resolve Ring~2 in Figure~\ref{fig1} visually. The reason that we cannot extract Ring 2 from STIS disk modeling is caused by the noise between Ring 1 and Ring 2 in Figure~\ref{fig1}c. We calculate an apparent rotation rate of $15\fdg5\pm9\fdg9$~yr$^{-1}$ clockwise between the two observations. Although we do not visually detect Ring~2 in the NICMOS data, the corresponding azimuthal profile can be fit with a cosine profile, and the dip in NICMOS is shifted ${\sim}20^\circ$ from that in SPHERE, see Figure~\ref{fig-ring2-dip-motion}a. 

The dip might have rotated for an additional $360^\circ$ between NICMOS and SPHERE that establish a $19$~yr baseline. Such an additional rotation is likely. When we include this rotation, we can explain the motion of the dip in the three instruments with a $19\fdg8\pm1\fdg0$~yr$^{-1}$ clockwise rotation in Figure~\ref{fig-ring2-dip-motion}b. Under this scenario, for a $0.46_{-0.10}^{+0.07}~M_\sun$ central star, the rotation of the dip traces a shadowing source that is under Keplerian rotation at $5.3_{-0.5}^{+0.4}$~au, or $0\farcs155_{-0\farcs015}^{+0\fdg012}$ that is just beyond SPHERE's coronagraph. Nevertheless, we reiterate that our NICMOS data reduction is prone to overfitting, and thus our measurement is not conclusive unless follow-up SPHERE observations are available: they would allow us to better investigate the motion mechanism of the Ring 2 dip (i.e., the northwest underdensity region of Figure~\ref{fig-ring2-residuals}a).
 
\subsubsection{Clump}\label{sec-clump}
We recover a southern clump that is located at ${\sim}3\arcsec$ from the star in all three instruments. Given that the proper motion of TWA~7 is $(\rm{pm_{RA}}, \rm{pm_{Dec}}) = (-118.75 \pm 0.02, -19.65 \pm0.03)$~mas~yr$^{-1}$ in \citet{gaiaedr3}, a background galaxy would have a relative motion of $(\Delta\rm{RA}, \Delta\rm{Dec}) = (2\farcs4772 \pm 0\farcs0004, 0\farcs4099 \pm0\farcs0006)$ with respect to TWA~7 between NICMOS and STIS observations, the clump is therefore not a background galaxy. In comparison with the ALMA millimeter observations in 2016 in \citet{bayo19}, the southern elongated $2\sigma$ structure in their Figure~4 (from ${\approx}3\farcs5$ to $5\farcs5$, with a position angle of $195^\circ$) is consistent with the STIS clump in Figure~\ref{fig3}c.

We have a clear detection of the ${\approx}150$~au$^2$ clump in STIS with a sensitivity of ${\sim}1~\mu$Jy~arcsec$^{-2}$. Although STIS best resolves the clump, it is expected to map dust particles that are smaller than those in NICMOS and SPHERE. We thus do not compare the STIS data with the other instruments for motion analysis; instead, we compare the NICMOS and SPHERE observations to estimate its motion.

The azimuthal motion of the clump between NICMOS and SPHERE is $3^\circ\pm1^\circ$ using cross-correlation, which corresponds to the Keplerian motion at $130_{-20}^{+50}$~au from a $0.46~M_\sun$ central star. The derived location is consistent with the location of the clump that starts at ${\sim}3\arcsec = 102$~au within $2\sigma$. In addition, there is a marginal evidence in Figure~\ref{fig-spiral-image} that the motion rate of the clump decreases as a function of stellocentric separation. Nevertheless, since the clump extends beyond the edge of the NICMOS detector, and the NMF data reduction method may have altered its surface brightness distribution in NICMOS, we do not perform a clump motion measurement as for the S1 spiral. In addition, we cannot properly constrain the radial motion of the clump if it is moving outwards.

In comparison, \citet{boccaletti15} report apparently outward motion of ripple-like features for the edge-on debris disk orbiting M1 star AU~Mic, and \citet{chiang17} suggest that these features are sub-micron dust particles that are repelled by stellar wind. Limited by the NICMOS field of view, data quality, and data reduction artifacts, we cannot properly constrain the apparent motion for the clump in TWA~7. Nevertheless, the clump in the TWA~7 system might be under the same mechanism as the ripples in AU~Mic. We fit Gaussian radial profiles to the surface density distribution of the clump in \textit{HST}, and obtain peak location of $116\pm4$~au and $130\pm5$~au for NICMOS and STIS, respectively. Assuming STIS and NICMOS map the same dust source, this offset corresponds to a radial speed of $3.2\pm1.4$~km~s$^{-1}$ outwards. In comparison, the escape speed is $1.8$~km~s$^{-1}$ at $130$~au for a $0.5~M_\sun$ star. The clump is thus possibly unbound to the star; yet we reiterate that our calculation is based on the multiple assumptions that can pose challenges to our motion measurement.

\subsection{Improving point source detection}\label{sec-contrast}
\begin{figure}[htb!]
	\centerfloat
	\includegraphics[width=0.5\textwidth]{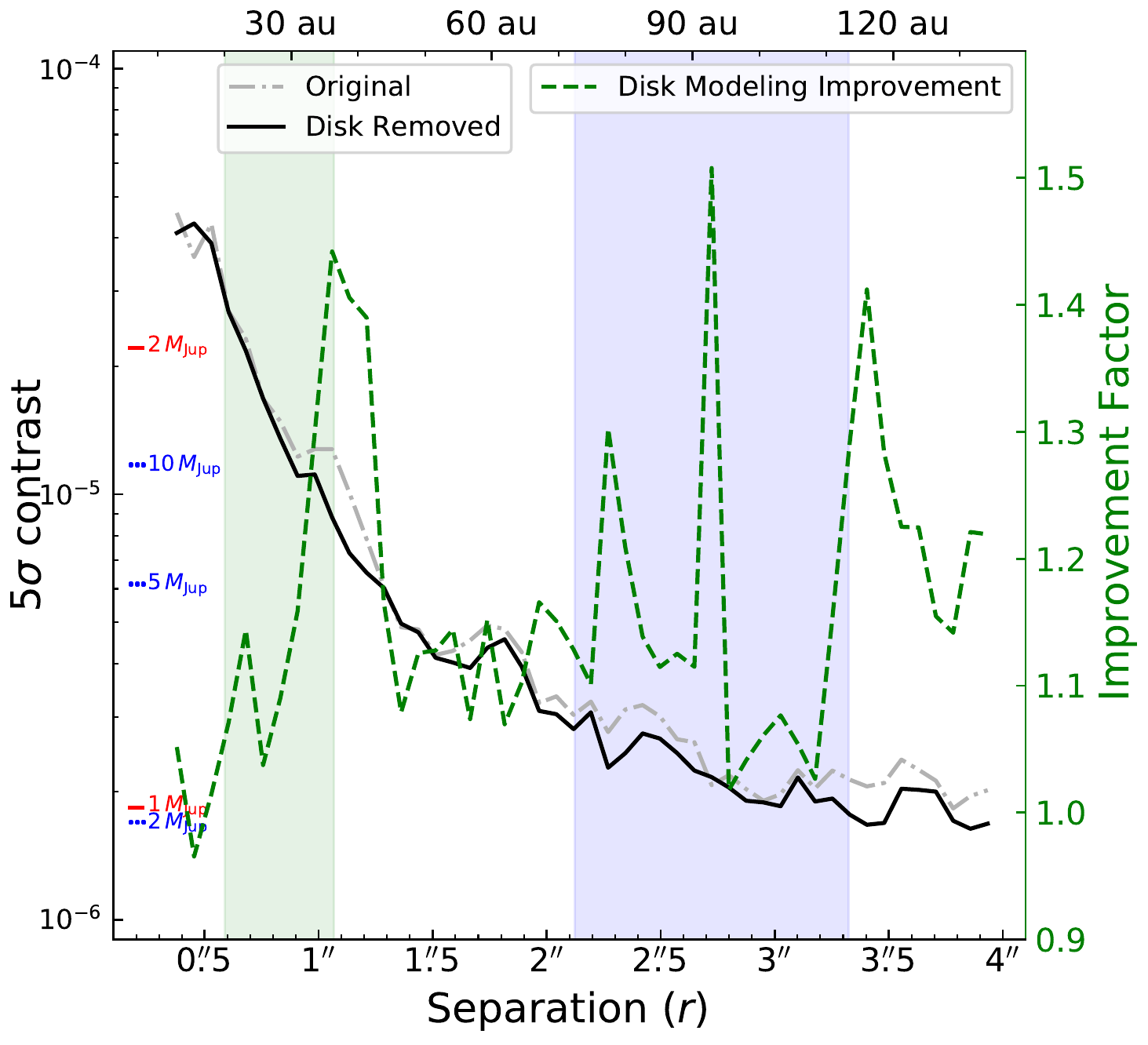}
    \caption{$5\sigma$ detection limit of point sources before and after disk removal for NICMOS, and improvement factor defined by dividing the two curves. The the full width at half maximum areas for the Ring 1 and Ring 3 models are shaded. For a $6.4$~Myr system, the mass of hot-start and cold start planets in \citet{spiegel12} are marked in red and blue, respectively. On the one hand, the disk removed contrast curve is deeper than the original one; on the other hand, the contrast improvement is the best near the location of Ring~1: both prove that disk modeling can help improve contrast.}
    \label{fig-nic-contrast}
\end{figure}

To investigate the impact of disk modeling on point source detection, we calculate for the NICMOS observation the $5\sigma$ contrast curves under two scenarios: before and after removing the best-fit model from the observations. We inject the {\tt TinyTim} PSF of TWA~7 to a physical on-sky position on the observations, perform reference differential imaging using KLIP (using the same number of references and components as NMF), then calculate the S/N for that position following the recipe in Section~3.4 of \citet{debes19}. For each location, we vary the flux of the injected point source through a binary search in log scale, until when its S/N is within 0.01 from 5, the corresponding planet-to-star flux ratio is then the $5\sigma$ detection limit. We define contrast as the flux of the detection limit divided by $1.296$~Jy (i.e., the NICMOS flux of TWA~7. See Section~\ref{sec-color} for the calculation.). For each on-sky  radial location, we perform this calculation with an azimuthal step of $30^\circ$, and calculate the mean contrast for all azimuthal positions. We repeat the above process for each radial separation between $5$ NICMOS pixel and $53$ NICMOS pixel (i.e., $0\farcs38$ to $4\farcs01$), and present the contrast curves in Figure~\ref{fig-nic-contrast}.

By dividing the two contrast curves, we notice that the disk-removed contrast curve is deeper than the original contrast curve by $m =16\%\pm12\%$. What is more, the contrast improvement can reach $m=45\%$ near the location of Ring~1, or $m\approx25\%$ near Ring~3.\footnote{For a small number $m$ (specifically, $|m| < 0.6$), we have $2.5\log_{10}(1+m)\approx\ln{10}\cdot\log_{10}(1+m)=\ln(1+m)\approx m$. The improvement in magnitude thus has a nearly identical value as the flux ratio in this study.} Our comparison of the two curves demonstrates the necessity of disk modeling in detecting fainter planets.

 \begin{figure*}[bht!]
	\includegraphics[width=\textwidth]{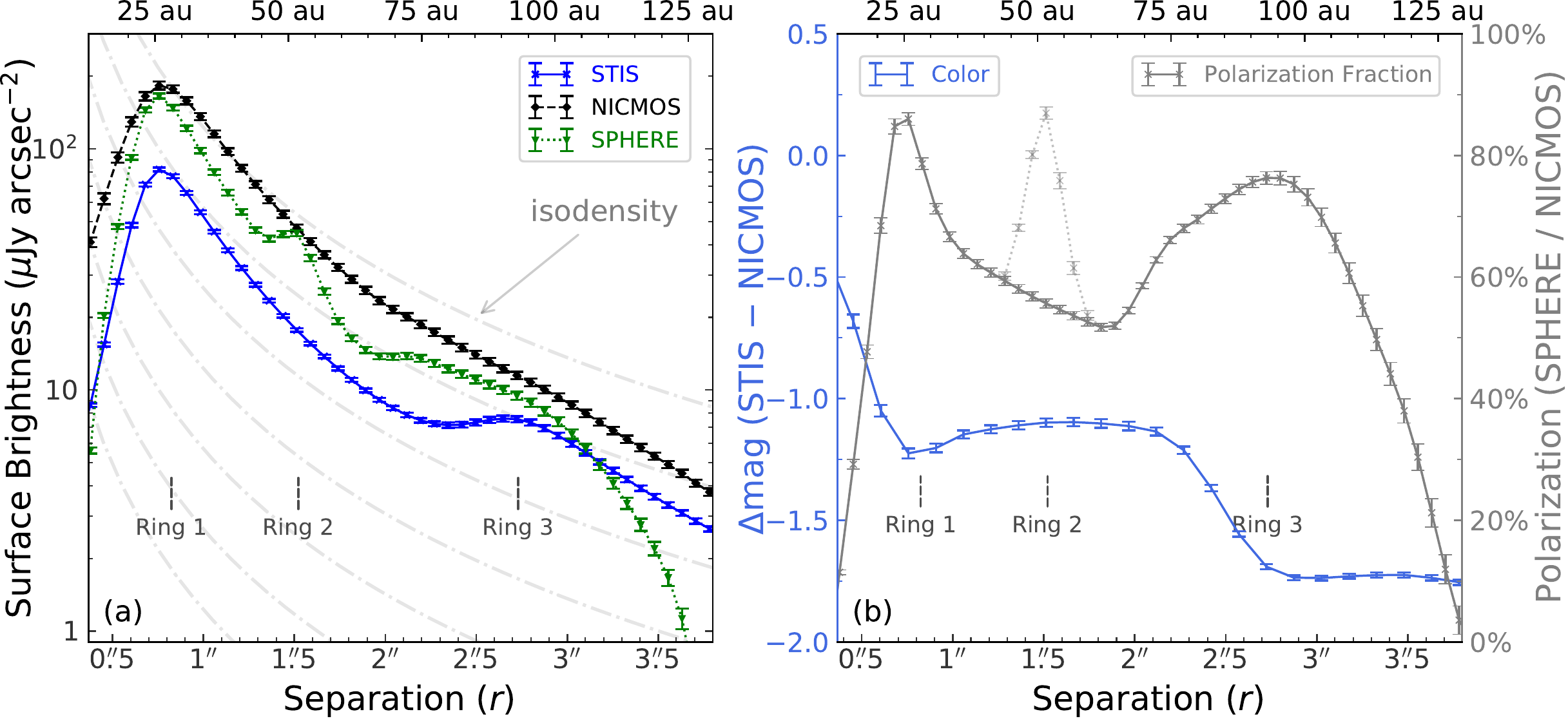}
    \caption{(\textbf{a}) Radial profiles measured on best-fit models. (\textbf{b}) Radial profiles for color and linear polarization fraction, measured from STIS$-$NICMOS after correction of instrumental response and SPHERE$/$NICMOS, respectively. Due to the non-detection of Ring~2 in NICMOS modeling, the polarization fraction for Ring~2 is expressed in dotted line. Note: the error bars in both panels are $3\sigma$ standard errors.
    }
    \label{fig-radial-color-pol}
\end{figure*}

We have assumed that there are no detectable planets in our contrast curve calculation for NICMOS observations. Were there hypothesized planets that should be detectable using the NICMOS data, such planets might have been overfit by the best-fit disk model. An ideal approach is to mask out different positions and perform disk modeling then contrast calculation, since such a procedure could extract these planets \citep[e.g., PDS~70c:][]{wang20} and consequently help detect fainter planets while enhancing the improvement in detection limit. We do not perform such a calculation, since we have not detected point sources with high confidence, and the presented results Figure~\ref{fig-nic-contrast} have demonstrated the necessity of disk modeling in detecting fainter planets.

\section{Dust Properties}\label{sec-dust-prop}

We study dust properties for the system using the best-fit models in the three instruments, see Figure~\ref{fig-radial-color-pol}a for the surface brightness radial profiles used for analysis in this Section. To compare the radial profiles, we have interpolated the best-fit images at a resolution of twice the NICMOS pixel size. 

In comparison with Figure~\ref{fig1}d, the recovered NICMOS surface brightness is at a level that is brighter than SPHERE, which confirms the expectation that NICMOS data reduction with NMF is under the influence of overfitting. We thus use the best-fit models to minimize such effects in comparing the system at different instruments. Nevertheless, we only recover the plateau at the Ring~2 location in the SPHERE model, we thus do not include the Ring~2 component in our analysis here.

\subsection{Color}\label{sec-color}
Using the best-fit NICMOS and STIS models, we calculate the color for the scatterers. To account for the differences such as stellar brightness at the two wavelengths and instrument response, we first calculate the unobstructed instrumental responses to the \citet{kurucz93} star model of TWA~7 using {\tt pysynphot}  \citep{pysynphot}. The {\tt pysynphot} inputs are the effective temperature for TWA~7 \citep[4018~K:][]{gaiadr2}, its $V$-band magnitude \citep[10.91:][]{2006AA...460..695T}, and its surface gravity $\log g$ in units of cm~s$^{-2}$ \citep[4.18:][]{tess_input_catalog}. For NICMOS F160W, the {\tt ObsBandpass} input parameter is {\tt `nicmos,2,f160w'}, and the corresponding instrument response is $0.953$~Jy (yet we adopt a value of $0.953\times\frac{2.034\times10^{-6}}{1.496\times10^{-6}}=1.296$~Jy to take into account the difference in the {\tt PHOTFNU} parameters in the two NICMOS eras); for STIS, {\tt `stis,ccd,a2d4'}, and $0.177$~Jy. Then for each instrument, we divide the surface brightness distribution by the corresponding instrumental response, and obtain the relative brightness between the disk and the star.  The integrated relative reflectance, $F_{\textrm{disk}}/F_{\textrm{star}}$, is $3.53\times10^{-3}$ in STIS, and $1.01\times10^{-3}$ in NICMOS.

We calculate the relative brightness radial profiles, then convert them to magnitude per square arcsec, and obtain the color radial profile by subtracting the converted NICMOS profile from the STIS one, see the results in Figure~\ref{fig-radial-color-pol}b. We observe that the relative reflectance in STIS is higher than that in NICMOS, making the scatterers blue. In addition, the system is bluer when stellocentric distance increases: Ring~1 is relatively brighter in STIS with $\Delta\textrm{mag}=-1.25$, while Ring~3 has $\Delta\textrm{mag}=-1.75$. Both phenomena are consistent with the expectations that M stars can retain small dust particles in their debris disks \citep[e.g.,][]{arnold19}, and that small particles which can scatter more light at shorter wavelengths are pushed further out than larger ones \citep[e.g.,][]{strubbe06, thebault08}.

\subsection{Polarization fraction}

We investigate the polarization properties for the scatterers since NICMOS and SPHERE cover similar wavelengths: the former probes the system in total intensity at $1.4$--$1.8~\mu$m, the latter in polarized light at $1.48$--$1.77~\mu$m. We calculate the linear polarization fraction by dividing the surface brightness profile of the best-fit SPHERE model by that of the best-fit NICMOS model, see Figure~\ref{fig-radial-color-pol}b. Given that we cannot recover Ring~2 with NICMOS modeling, we only focus on Ring~1 and Ring~3 here. 

The Ring~1 region has a peak polarization fraction of $85\%$, while the Ring~3 region peaks at $75\%$. High polarization is consistent with the explanation that the scatterers are small and/or not compact: on the one hand, high polarization is expected theoretically for small dust particles ($0.1~\micron$ -- $0.2~\micron$: \citealp{dabrowska13, perrin15}); on the other hand, similar levels of polarization fraction have been experimentally measured for fluffy aggregates \citep{volten07}. Given that there could exist non-negligible stellar wind activity around M stars, and that we cannot constrain the activity for TWA~7 using existing indicators such as X-ray brightness (Section~\ref{sec-intro-twa7}) and Ring 1 tail distribution (Section~\ref{sec-ring1-tail}), we do caution that small dust particles created by collisional cascade can be efficiently removed by stellar winds.

\subsection{Ring~1: Rayleigh scattering in total intensity?}\label{sec-ring1-rayleigh-or-not}
\begin{figure}[htb!]
	\centerfloat
	\includegraphics[width=0.48\textwidth]{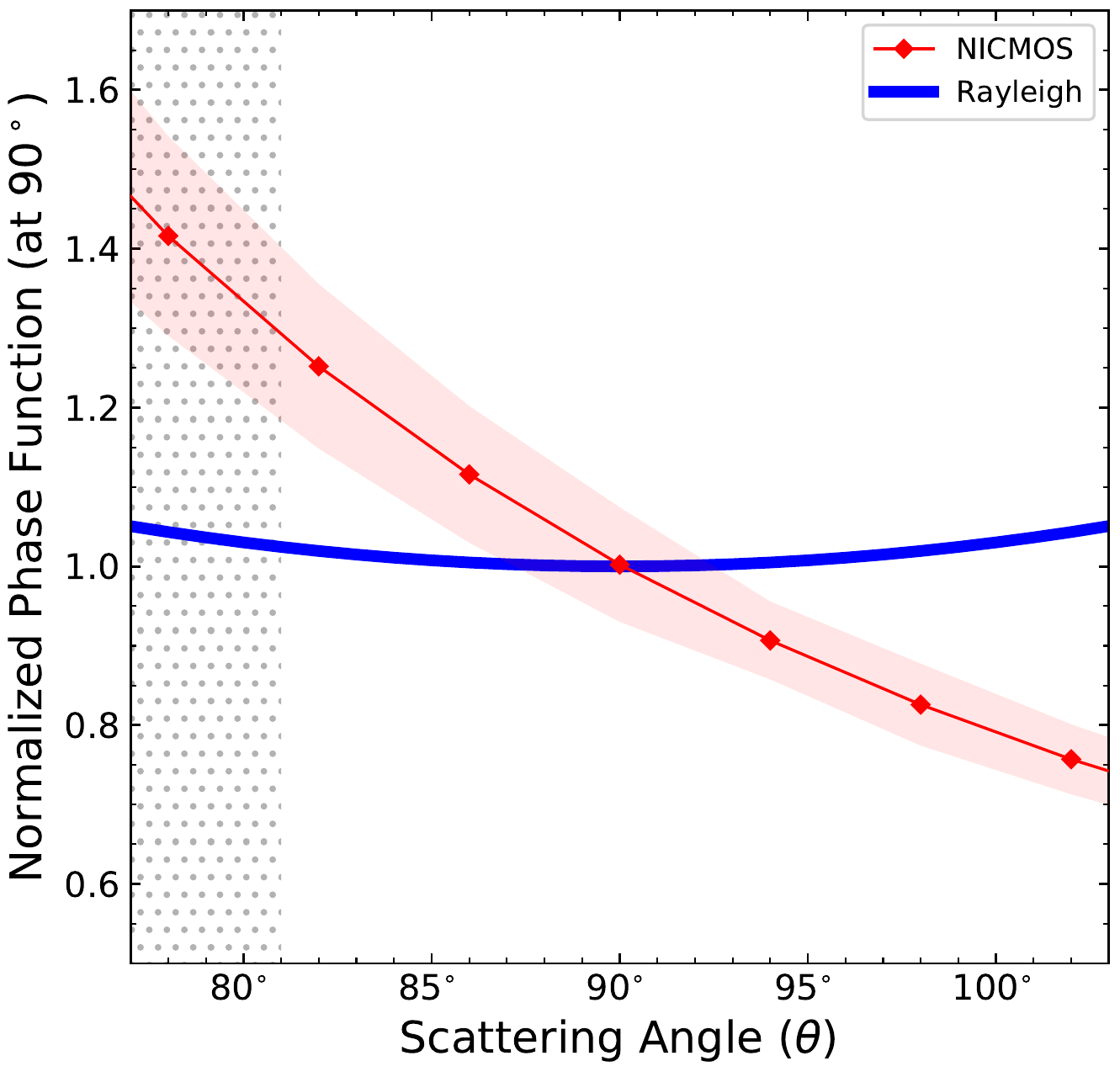}
    \caption{Normalized total intensity phase functions between TWA~7 Ring~1 in NICMOS and Rayleigh scattering show little overlap. However, we cannot use the parametric Henyey--Greenstein phase function to depict Rayleigh scattering, see Section~\ref{sec-ring1-rayleigh-or-not}. Note: the $77^\circ$--$81^\circ$ region shaded with dots is contaminated by S1.}
    \label{fig-rayleigh}
\end{figure}

In the Rayleigh scattering regime (i.e., the size of the scatterers is smaller than the observation wavelength by more than one order of magnitude), the scattering phase function in total intensity follows
\begin{equation}
I_\textrm{tot}(\theta) = \frac{3}{4}\left(1+\cos^2\theta\right),
\end{equation}
while the polarization fraction follows Equation~\eqref{eq-ray-mod}. We sample $5000$ data points from the NICMOS phase functions from Ring~1 modeling for comparison, see Figure~\ref{fig-rayleigh}. The NICMOS phase functions do not have a large overlap with the Rayleigh phase function in total intensity. Nevertheless, we cannot conclude on the scattering regime for the TWA~7 disk seen in  NICMOS. 

Mathematically, the parametric Henyey--Greenstein phase function in Equation~\eqref{eq-hg} is a monotonic function for $0^\circ \leq \theta \leq 180^\circ$, while the Rayleigh total intensity phase function is symmetric about $\theta = 90^\circ$ that cannot be depicted by the Henyey--Greenstein function. Therefore, we cannot test whether a system is under Rayleigh scattering in NICMOS total intensity using Henyey--Greenstein phase functions. Similarly, we cannot test whether the SPHERE data in polarized light is under the Rayleigh scattering regime with the modeling presented in this paper. 

Physically, collisional cascade produces significantly more small dust particles than larger ones, i.e., $n(a) \propto a^{-3.5}$ where $a$ is the particle size \citep[e.g.,][]{dohnanyi69}. The size of the particles in debris disks can reach the observation wavelength in scattered light. The existence of particles with sizes comparable with observation wavelength makes it necessary to model them using Mie theory and its variations, or more complicated radiative transfer modeling methods. 
 
To determine the scattering regime for Ring~1, especially to test the $\lambda^{-4}$ dependence of scattered light intensity as a function of wavelength $\lambda$, future high resolution multi-wavelength observations of TWA~7 in total intensity, as well as the actual Rayleigh scattering phase function, are needed.

\section{Summary}\label{sec-sum}
From a favorable nearly face-on vantage point of the M star TWA~7 system, we have performed  a multi-instrument and multi-decade characterization of its circumstellar disk in scattered light. By analyzing the TWA~7 observations with \textit{HST}/STIS,  \textit{HST}/NICMOS, and VLT/SPHERE using disk modeling, we report a layered debris structure around M star TWA~7. We identify three rings (Ring 1, Ring 2, and Ring 3), one spiral (S1), and an ${\approx}150$~au$^2$ dusty clump. We also tentatively detect a secondary spiral (S2) that has low average pixelwise S/N. Our disk modeling using static geometric models shows that the extended Ring 1, which peaks at $0\farcs8$, dominates the overall distribution of dust. The rest of disk components are superimposed onto the outskirts of the extended ring.

M stars have stellar winds bring corpuscular and Poynting--Robertson drag, and consequently their debris disk tail have a surface density power law index from $-1.5$ to $-2.5$ depending on the stellar mass loss rate $\dot{M}_\textrm{star}$ \citep{strubbe06}. However, if Ring 1 is the birth ring of small particles (see \citealp{pawellek19} for exceptions), the power law index of $-0.7$ for its tail is outside the expected range: it is even shallower than that for quiescent stars with $\dot{M}_\textrm{star} \lesssim 10\dot{M}_\Sun$. The existence of secondary CO gas in this system might help explain the power law index for Ring 1. With additional complicating factors such as the existence of Ring 2 and Ring 3, we cannot infer the stellar wind activity for TWA~7 with our power law measurements. Nevertheless, such constraints might be obtained by fitting the multi-instrument observations and retrieve dust properties such as size and porosity, or by a dynamically motivated study of the spatial distribution of the system \citep[e.g.,][]{schueppler15}.

We have compared the reference differential imaging detection limit of point sources for NICMOS before and after disk modeling.  By modeling the NICMOS disk, we can detect planets that are fainter by ${\sim}15\%$ overall. What is more, at the Ring~1 location where the disk is the brightest, we can improve the detection limit by $45\%$. Our study demonstrates the necessity of disk modeling in detecting fainter planets for upcoming high-contrast imaging missions, e.g., \textit{Roman}/CGI, \textit{LUVOIR}, and \textit{HabEx}.

We have compared the surface brightness distribution of the system in three instruments. On the one hand, using NICMOS and STIS total intensity images, we observe that the system has a blue color. The system is bluer when the stellocentric distance increases. 
On the other hand, using SPHERE and NICMOS images, we compare the system in polarized light and total intensity: ${\sim}80\%$ of the light is polarized. The former is consistent with the expectation that small particles scatter light more efficiently at shorter wavelengths, the latter is consistent with the expectation for small or fluffy dust. Nevertheless, we caution that the existence of stellar winds can change our inference. In addition, by comparing the total intensity phase functions between NICMOS and Rayleigh scattering, we find that the Henyey--Greenstein approach cannot be used to test the Rayleigh scattering regime.

We have calculated the rotation of the S1 spiral arm, and the radial motion of the clump,  after removing the disk models from the observations. The S1 rotation can be described by both solid body movement and local Keplerian movement. The radial motion of the clump is possibly unbound to the system. Although the observations establish a two-decade timeline for motion analysis, multiple factors including the data reduction method, pixel size difference, stellar alignment, and different field of view of the detectors, keep us from determining the motion mechanisms for S1 and the clump.

We have inspected the morphology and apparent rotation for Ring~2, and investigated two possible scenarios for its formation that can be tested with new SPHERE observations. On the one hand, the optical depth distribution of Ring~2 in SPHERE might result from mean motion resonances with a planetary perturber that is located at the northwest underdensity region. Were the system under this scenario, the Ring~2 and Ring~3 components, which are spatially separated in our disk modeling, would be physically produced simultaneously by the hidden perturber. On the other hand, the azimuthal motion of an underdensity region (or a brightness dip) on Ring~2 can be explained by shadowing effects from a Keplerian rotating disk at $5.3_{-0.5}^{+0.4}$~au.

The layered debris disk around TWA~7 in scattered light reveals that M stars can host complex debris structures, and the motion of these structures can help understand their formation mechanisms. SPHERE re-observations after 2021 will establish a ${>}4$ year timeline to help constrain the motion for the S1 spiral and the clump, investigate the existence of the S2 spiral, and examine the morphology and motion for Ring~2. Being debris disks orbiting M stars, the almost face-on TWA~7 system offers a complementary view than the edge-on AU~Mic system. Being members of the TW~Hya association, the existence of the face-on debris disk around TWA~7, and the face-on protoplanetary disk around M star TW~Hya (i.e., TWA~1), shows that circumstellar disks can evolve at different rates even for similar spectral types.

\acknowledgements
We thank the anonymous referee for their useful suggestions that make this paper more thorough. We thank Johan Olofsson and Rob van Holstein for discussion on SPHERE/IRDIS data reduction and calibration. Based on observations made with the NASA/ESA \textit{Hubble Space Telescope}, obtained from the data archive at the Space Telescope Science Institute. STScI is operated by the Association of Universities for Research in Astronomy, Inc.~under NASA contract NAS 5-26555. This research has made use of data reprocessed as part of the ALICE program, which was supported by NASA through grants \textit{HST}-AR-12652 (PI: R. Soummer), \textit{HST}-GO-11136 (PI: D.~Golimowski), \textit{HST}-GO-13855 (PI: E.~Choquet), \textit{HST}-GO-13331 (PI: L.~Pueyo), and STScI Director's Discretionary Research funds, and was conducted at STScI which is operated by AURA under NASA contract NAS5-26555. The input images to ALICE processing are from the recalibrated NICMOS data products produced by the Legacy Archive project, ``A Legacy Archive PSF Library And Circumstellar Environments (LAPLACE) Investigation,'' (\textit{HST}-AR-11279, PI: G.~Schneider). Based on observations collected at the European Organisation for Astronomical Research in the Southern Hemisphere under ESO programme 198.C-0209(F). Part of the computations presented here were conducted on the Caltech High Performance Cluster, partially supported by a grant from the Gordon and Betty Moore Foundation.  

\facilities{\textit{HST}  (NICMOS, STIS), VLT:Melipal (SPHERE)}

\software{{\tt DebrisDiskFM} \citep{ren19}, {\tt emcee} \citep{emcee}, {\tt IRDAP} \citep{vanholstein20}, {\tt nmf\_imaging} \citep{nmfimaging20}, {\tt pysynphot}  \citep{pysynphot}, {\tt TinyTim} \citep{tinytim}}

\bibliography{refs}
\end{CJK*}
\end{document}